\documentclass[11pt]{article}

\usepackage{amsmath}
\usepackage[breaklinks=true,
              hypertexnames=false,
              linktocpage,
              plainpages=false,
              pdfpagelabels]{hyperref} 
\numberwithin{equation}{section}
\usepackage{cleveref}
\usepackage{fullpage,amscd,amsthm,upref,graphics}
\usepackage{titling} % Customizing the title section
\usepackage{authblk}
\usepackage[T1]{fontenc}
\usepackage[utf8]{inputenc}
\usepackage{slashed}

\usepackage{amstext}
\usepackage{amsfonts}
\usepackage{bbold}
\usepackage{nccmath}
\usepackage[export]{adjustbox}
\usepackage{soul}
\usepackage{microtype}
\usepackage{cite}
\usepackage{enumerate}
\usepackage[english]{babel}
\usepackage{epsfig,amssymb}
\usepackage{dcolumn,color}
\usepackage{bm}
\usepackage{graphicx}
\usepackage{footnote}
\usepackage{float}

\usepackage{xcolor}
\usepackage{lmodern}
\newcommand{\eqcomma}{,\phantom{AA}}

\numberwithin{equation}{section} \numberwithin{table}{section}
\numberwithin{figure}{section}

\title{Linear response hydrodynamics of a relativistic dissipative fluid with spin}
\author{Giorgio Torrieri$^1$, David Montenegro$^2$ \\
\emph{\small{$^1$ IFGW, Unicamp, Campinas, Brasil}}\\
\emph{\small{$^2$Bogoliubov Laboratory of Theoretical Physics, JINR, 141980 Dubna, Moscow region, Russia.}}\\}

\begin{document}
\maketitle

\begin{abstract}
 We formulate a lagrangian hydrodynamics including shear and bulk viscosity in the presence of spin density, and investigate it using the linear response functional formalism.    The result is a careful accounting of all sound and vortex interactions close to local equilibrium.
   
\end{abstract}

\newpage
 
\tableofcontents

\section{Introduction}

Relativistic dissipative hydrodynamics has been part of a vigorous theoretical investigation, triggered by the experimental discovery of a vorticity-correlated hyperon spin polarization and vector resonance spin alignment in heavy ion collisions \cite{lisa}. Recently, several versions of hydrodynamics with spin have been proposed  \cite{dm1,dm2,dm3,flork1,hydroryb,flork2,flork3,gale,hongo,bec,shearspin,nair,dirk,kam,palermo} but basic conceptual questions, such as the role of spin-vorticity coupling, pseudogauge dependence, entropy production and the definition of equilibrium, remain unanswered.
 
One approach that has the advantage of an immediate connection with both microscopic statistical mechanics and field theory is Lagrangian hydrodynamics \cite{hydroryb} analyzed via linear response techniques.   In this formalism, one abandons the definitionof hydrodynamics as dictated by {\em conservation laws}, but instead develops it from a definition of {\em a free energy} to be locally minimized.   The advantages, as written earlier, are a direct connection with microscopic entropy and, since conservation laws do not dictate the dynamics, a bypassing of the pseudo-gauge issue.  The disadvantage is that away from the ideal limit free energy is not maximized {\em exactly} so a Schwinger-Keldysh formalism, and a precise fluctuation-dissipation relation, are needed to hold for the dynamics to be well-defined\cite{gavassino}.

The lagrangian approach was previously used to understand the properties of a non-dissipative fluid \cite{hydro0,hydro1,hydro2}, then one with bulk \cite{groz} and shear \cite{uslag} viscosity as well as Israel-Stewart ($IS$) corrections.
It was used in \cite{dm1,dm2} to understand the transport properties of a nearly ideal fluid with spin and to show that this limit clashed with causality \cite{dm3}.  Finally, \cite{DM} a rigorous connection was made between spin hydrodynamics and the linear response theory of \cite{kadanoff,tong}, including a Higgs-like relation between the polarization ``condensate'' and spacetime symmetries and a fluctuation dissipation relation between polarization susceptibility and spin relaxation time, whose long-time tail parallels the tail of hydrodynamic fluctuations \cite{kovtail}.

In this work, we would like to extend our earlier analysis \cite{DM} to include dissipative effects as well as
spin effects.   This is because, while spin is a necessary contributor to dissipation \cite{dm3}, it is not its only source, since of course momentum diffusion by microscopic collisions is still present.   As recently seen in \cite{kam}, the interplay of the ``pure dissipation'' scale with the ``Polarization scale'' can be quite non-trivial (as argued later the polarization scale is the first inequality of \eqref{hyerarchy} which however is not necessarily negligible).
Moreover the broken time-reversal symmetry induced by dissipation, and the broken spacetime symmetry induced by polarization, can combine in relatively non-trivial ways.    In this work we will use the linear response analysis developed in \cite{DM} to elucidate all these issues in detail, clarify how polarization and dissipation combine in different regimes, and explain the role of interactions between ``macroscopic'' sound and diffusive modes and microscopic fluctuations as well as of the effective symmetries. In \cref{deux}, we review the QFT implementation via the Schwinger-Keldysh formalism and the construction of both dissipative and spin hydrodynamics as effective field theories. In \cref{chap:3}, we will put together the dissipative and polarization terms to study the interaction between them. This analysis is developed via Feynman diagrams in \cref{chap:4} by deriving the EoS and transport corrections due to polarization and elucidating the regimes in the different length scale hyerarchies. In \cref{mef} we provide an overview of the Markovian limit and discuss the inclusion of the memory effect to describe second-order polarized hydrodynamics. Finally, in \cref{sao} we present the general remarks and summarize our findings. The detailed form of the Kubo formulae and correlations are given in \cref{gk} and \cref{twop}. Finally, we address the memory effect by means of hierarchy of the relaxation time scale. 

\vspace{1em}

\noindent {\it Conventions:} $\hbar=1$. $c=1$ 
metric $\eta_{\mu\nu}=\text{diag}(+,-,-,-)$.  $\mu,\nu,\dots$ all space-time coordinates,  $i,j,\dots$  spatial coordinates.

\section{Schwinger-Keldysh \texorpdfstring{$CTP$}{CTP} formalism}\label{deux}

We have known hydrodynamic as a nonlinear effective description in the infrared regime where the properties occur from underlying microscopic interaction. In this sense, it serves as a guide to theoretical progress that "bottom-up", instead of "top-down", can be used to capture phenomenological aspects. Our approach is a good qualitative description of what is going on in a dissipative polarized fluid, and the justification of this method will follow when we get the analytical results. In this section, we use the closed time path formalism to study non-equilibrium field theory for the effective field theory of a dissipative polarized fluid in rotation. We claim to extract useful information to understand this phenomenon from a better theoretical viewpoint. For simplicity, we are primarily interested in a near-equilibrium phase transition whose long-wavelength fluctuations dominate correlation functions.

\subsection{\texorpdfstring{$CTP$}{CTP} Wilsonian effective action}\label{noneq}

The Schwinger-Keldysh formalism has been developed to exploit the fluctuation-dissipation theorem to calculate deviations from equilibrium. These deviations are modeled as perturbations on a heath bath. The generating functional 
\begin{equation}\label{first:eq}
Z[J] =  \int \mathcal{D} \varphi  \rho[\varphi (t,\textbf{x})] \exp \bigg\{ i S [\varphi] + i \int_\mathcal{C} d^4x J \cdot \varphi \bigg\},
\end{equation} 
where $\varphi$ encodes $\{ \phi, \Psi \}$ the infrared and ultraviolet degrees of freedom, respectively, $J$ is the usual classical external source, and the $\rho$ is the density matrix. The integration contour $\mathcal{C} = \mathcal{C}_1 \cup \mathcal{C}_2 $ represents the branch $\mathcal{C}_1$ running "towards" positive time axis and $\mathcal{C}_2$ running "backward" negative time axis. 
% To illustrate the general effective field theory idea, the $\phi$ .  We can write the 
The $n-$point correlation function is
\begin{equation}\label{old:green2}
G (x_1,..,x_m) = i (-1)^{m} \frac{\delta^m \ Z [J]}{\partial J (x_1) \dots  \partial J (x_m)}  = \langle \mathcal{T} \{ \phi (x_1) \ldots \phi (x_m) \} \rangle,     
\end{equation}
where $\mathcal{T}$ is the time-ordered-product and $\langle \ldots \rangle$ the thermal expectation value\footnote{The coordinate arguments is in 4-dimensional Minkowski space.}. Since the equilibrium is built around stationarity deviations, implemented via the $KMS$ condition. It must necessarily come in two types, "towards" and "away" from equilibrium. Microscopically, the equilibrium condition demands that transitions to and from equilibrium must be equivalent, but in the long run, as equilibrium maximizes the number of microstates (whose logarithm is the entropy), fluctuations towards equilibrium will be more frequent. Within a field theory defined via functional integrals, these requirements are satisfied by doubling of degrees of freedom 
\begin{align}
\varphi^+ = \frac{\varphi_1 + \varphi_2}{2}, \qquad \varphi^- = \varphi_1 - \varphi_2, \qquad J^+ = \frac{J_1 + J_2}{2}, \qquad J^- = J_1 - J_2,
\end{align}
as well as the Hilbert space with the purpose to approach any real process in the effective theory. Substituting these coordinates in \eqref{first:eq}, the generator functional in the Heisenberg representation becomes 
\begin{equation}
Z [\varphi_\pm , J_{\pm}] = e^{iW[\varphi_\pm, J_\pm]} = Tr [U (+ \infty,- \infty,\varphi_+,J_+) \rho_o [\varphi_\pm (t,\textbf{x})] U^\dagger (+ \infty,- \infty,\varphi_-,J_-)],
\end{equation}
being $\rho_o  $ the initial density matrix. This transformation explicitly makes the connection between the ``exact'' quantum interaction with a thermal bath. As the unitary matrix is $U= \mathcal{T}\{\exp{i\int d^4 x [\mathcal{E}(x)- J^+ (x)\varphi^+ (x)]}\}$, the generator function can be decomposed as
\begin{equation}\label{old:Z}
Z [\phi_\pm, \Psi_\pm] =  \int \mathcal{D} \phi_\pm  \mathcal{D}  \Psi_\pm  \rho [ \phi_\pm , \Psi_\pm ]  \exp \bigg[ i S [\phi_\pm] + i S [ \Psi_\pm] + i S [\phi_\pm, \Psi_\pm]  \bigg],  
\end{equation}
where the conservative canonical stress-tensor of  $S [\phi_\pm]$ is bounded from below, and $S[\Psi_\pm]$ is microscopic variables action. The fact that there are two directions, "towards" and "away" from equilibrium, means that, from a microscopic viewpoint, the coarse-grained procedure provides both the fluctuation and dissipation in $S [\phi_\pm, \Psi_\pm]$. Moreover, it is misleading to interpret the effective field theory as an infrared limit of an old-fashioned theory whose dynamics of microscopic variables are unknown or impractical to describe. The effective field theory is built out of \textit{relevant} macroscopic fields to investigate the problem. It entails that even though we cannot keep track of inaccessible degrees of freedom, their feedback interferes with the dynamics of accessible degrees of freedom. 

After performing the integration out of the initial conditions and microscopic variables ($\Psi$-sector) from \eqref{old:Z}, we have
\begin{equation}
Z [\phi_+, \phi_-] = \text{Const} \times \int \mathcal{D} \phi_\pm \rho [ \phi_\pm ] \exp{ [ \underbrace{ i S [\phi_\pm] + i \lambda \int d^4 x  \phi^a \hat{G}_{ab} \phi^b 
}_{S_{CTP}}]} ,  
\end{equation}
where the density matrix $\rho [\Psi_\pm]$ is integrated over. We omit the derivatives upon $\phi$ to keep the notation light. For sake of simplicity, we are interested in linearized perturbation, so the Green-Kubo formulae can play an important role to account for the time evolution of fluctuations, [see Eq. \eqref{stctpform}]. Note that all non-trivial couplings between $\phi-\Psi$ will be required according to the dissipative-order needed to exploit the fluid behavior. Summarizing the expression for the effective action as
\begin{equation}\label{mix::ctp}
S_{CTP} [\phi ^+,\phi ^-, J^+, J^-] = T^4_o \int_{t_i}^{t_f} d^4 x  \{ \mathcal{L}_o [\phi^+, J^+] - \mathcal{L}_o^* [\phi^-,J^-] + \mathcal{L}_s [\phi_\pm , J_\pm] \},
\end{equation}
where $t_i$ and $t_f$ are the initial and final time, respectively, and $T^{-1}_o $ is a fundamental energy scale where thermal fluctuations are relevant. It is useful, in this language, to separate the contributions coming from separate scales. The two copies of the non-dissipative lagrangian $\mathcal{L}_o$ and $\mathcal{L}^*_o$, for each doubled variable, manifest the conservation of variables comparable to the system dimension. The inclusion of $\mathcal{L}_s$ expresses the system-environment correlation, which consists of accessible-inaccessible degrees of freedom. More precisely, this lagrangian keeps the infrared dynamics almost closed and enlarges the coupling space with thermal noise and thermodynamic forces that, eventually, break symmetries. The action \eqref{mix::ctp} is subject to the symmetry $S_{CTP}[\phi^+,\phi^-]=-S^*_{CTP}[\phi^-,\phi^+]$. Expecting this restriction for non-dissipative process, the time-inversion is preserved through $S_{CTP}[\phi^+,\phi^-] =S_{CTP} [\phi^+,\phi^+] - S_{CTP} [\phi^-,\phi^-] $. However, in dissipative situations, one requires the imaginary part be positive $Im S_{eff} [\phi^+,\phi^-] >0$. As the boundary condition $\phi^+ (t_f) = \phi^- (t_f)$ introduces the reminiscent long fluctuation, we can still interpret the effective lagrangian as a gradient expansion of infrared-variables. In addition, we demand $ \frac{\delta^2 \mathcal{L}_s  }{\delta \phi^+ \delta \phi^- } \ne 0 $ since the collision of microscopic degrees of freedom and interactions with the external fields are regulated by the second derivative of the action. Including these fluctuations, a satisfactory definition of the Green Function of \eqref{old:green2} is easy to develop. By generating functional which describes the dynamical direction in $CTP$ internal space $\{ +,- \}$, one can manufacture a $2 \times 2$ matrix  
\begin{align}\label{stctpform}
\hat G_{ab} =\begin{pmatrix}G_{++}&G_{-+}\cr G_{+-}&G_{--}\end{pmatrix} \eqcomma  G_{mn} (x,x') = \bigg[ \frac{\delta}{i \delta J_m (x)} \bigg] \bigg[ \frac{\delta}{i \delta J_n (x')} \bigg] e^{i W [J_+,J_-]} \bigg|_{J_+ = J_- = 0},
\end{align}
where the generating functional to connected diagrams by $i W(J(x)) \equiv  \ln{Z(J(x))}$. The Green functions satisfy the algebraic identity $G_{++} + G_{--} = G_{+-} + G_{-+}$. Following the restrictions $\delta J_+ = \delta J_- = 0$, we define the classical fields as $ \phi_c (x) = \frac{\delta W[ \hat{\phi}_\pm ,\Omega_\pm] }{\delta J (x)}$. For arbitrary constant $a \in \mathbb{R}$, we have $ W[  \hat{\phi}_\pm  ,J_\pm] =  W[  \hat{\phi}_\pm + a ,J_\pm] $ with reflective condition: $ W[  \phi_+ , J_+ ; \phi_- , J_- ] = W[  \phi_- , J_- ; \phi_+ , J_+]$, and normalization condition:  $ W[  \hat{\phi} , J ; \hat{\phi} , J ] = 0 $. It is appropriate to define the generating functional of amputated $1PI$ by the functional Legendre transformation  
\begin{equation}
\Gamma [ \phi_+, \phi_-] =  W [ \hat{\phi}_\pm ,J_\pm] - c^{ab} \int_\mathcal{V} d^4 x \ J_a (x) \phi_b (x).
\end{equation}
It is clear that for a "classical" system one can use the method of steepest descent to obtain an estimate for $\Gamma$. In this limit, one has
\begin{equation}\label{gammadef2}
\left . \frac{\delta \Gamma}{\delta {\phi}_+(x)}\right|_{{\phi}_+={\phi}_-= \langle \phi \rangle}=0 \;,
\end{equation}
in analogy with statistical mechanics, where $\Gamma$ would represent free energy. When \eqref{gammadef2} has ill-defined or multiple minima, it indicates a phase structure with the Landau theory of phase transitions.  The $CTP$ formalism then allows calculating dynamical behavior (transport coefficients) around the phase transition \cite{hohenberg}, via the expansion around equilibrium outlined below, of course, the semi-classical limit neglects fluctuations. If the method of steepest descent is a good approximation, these can be modeled by an expansion of \eqref{gammadef2} around the minimum provided the effective degrees of freedom do not change, one can get the leading order propagator term 
\begin{equation}\label{propagator1}
\Gamma [\phi_+, \phi_-] = S_{CTP}[\phi_+,\phi_- ] - \frac{i \hbar}{2} \ln{\det \bigg( \frac{\delta^2 S_{CTP}}{\delta \phi_a (x) \delta \phi_b (x')}  \bigg)^{-1} [\phi_+,\phi_- ] } + \ldots.
\end{equation}
As we will show later, using the generating function technique, we get a self-consistent description of fluid and a general perspective of what symmetries are involved in dissipative phenomena. 

\subsection{Fluid dynamics as effective field theory} 

This section will be devoted to providing a big picture behind the generation of dissipative polarization by effective field theory technique. It will not be the first, many methods have been formulated to archive such variational description \cite{jeon}. However, these scalar theories present wrong calculations for one-loop diagrams because they fail to use the right symmetries, degrees of freedom, and interactions, so they end up describing particle subjects to a whatever potential instead of a perfect fluid.

As we have known, hydro is a nonlinear effective theory of long-wavelength that encodes collective motion of micro-particles. In the context of lagrangian hydrodynamics (actually any kind of continuum matter), the "field'' is nothing else than the position of a fluid cell, $\phi_{i=1,2,3}$. What distinguishes the ideal spinless fluid limit \cite{hydro0} are the reparametrization symmetries of these coordinates. As we will show later, using the generating function technique, we  get a self-consistent description of fluid and general perspective of what symmetries are involved in dissipative phenomena. 
\begin{subequations}\begin{align} 
\phi^{iI} & \to\phi^{iI}+\alpha^i, ~~ \text{with}~ \alpha^i=const., \\  
\phi^{iI} & \to R^i_j \phi^{jJ}, ~~\text{with}~ R^i_j \in SO(3), \\ \label{sy:3}
\phi^{iI} & \to\xi^{iI}(\phi), ~~\text{with}~ \det \bigg[  \partial \xi^{iI} / \partial \phi^{jJ}  \bigg]= 1,
\end{align}\end{subequations}
as well as the global infinite-dimensional deformation\footnote{Rewriting the $CTP$ internal space as the capital Latin indices $ \in \{ 0, 3 \}$, while the small ones, the spatial directions, run $1$ to $3$.}. The symmetry \eqref{sy:3} forces the lagrangian of ideal fluid to be of the form
\begin{equation}\label{ideal:}
\mathcal{L}_{free} = f (b), \ \ \ \  b=\sqrt{\det_{ij} \left[B_{ij}\right]} \eqcomma B_{ij}=\partial_\mu \phi_i \partial^\mu \phi_j.
\end{equation}
This variational principle deduces Euler's equations, but the converse is not generally true\footnote{It is worth pointing out the ground state is aligned with Euler coordinate frame, in this sense, the lowest energy state is a barotropic fluid which break translational and rotational invariance but not diffeomorphism.}. For further convenience, we shall replace the old-hydrodynamics variables with more suitable ones. In this case, we  identify the hydrodynamic entropy flow as a gradient expansion 
\begin{equation} 
K^{\mu I} \equiv P^{\mu\gamma I}_k \partial_\gamma \phi^{kK} \eqcomma 
P^{\mu \gamma I}_k =\frac{1}{6} \epsilon^{\mu \alpha \beta \gamma } \epsilon_{ijk}  \partial_\alpha \phi^{ i I } \partial_\beta \phi^{ j J}, 
\end{equation}
where $P^{\mu \gamma I}_K$ is a projector and ${ ( I , J ) = \{ (0,0) , (3,3) \}}$. The conservation law of entropy \cite{hydro2} is
\begin{align}\label{Kconser}
\partial_\mu K^{\mu I} = \partial_\mu \left(b u^{\mu I} \right)=0,
\end{align} 
where the $\phi^{iI}$ is invariant along its comoving frame $u^\mu\partial_\mu\phi^{iI}  = 0$. The velocity vector norm is $u^\mu u_\mu = 1 \rightarrow b^2 =  K^\mu K_\mu$. Note that the comoving projector is perpendicular to the flow direction 
\begin{equation}
\Delta^{\mu \nu} = B_{ij}^{-1} \partial^\mu \phi^{i} \partial^\nu \phi^{j} = \eta^{\mu\nu} - u^\mu u^\nu.
\end{equation}
It is easy to see $ u_\mu \Delta^{\mu\nu} = 0$. In the standard Hydrodynamic theory, the vortices $\omega^{\mu\nu}$ are the circulation of fluid velocity along a closed path on the singly connected region. One should translate them to the language of effective field theory. To do this, we realize that Kelvin's theorem can be written as an invariance under diffeomorphism volume $S\text{Diff}\,(\mathbb{R}^{1,3})$. Following this idea, the vortices can be phrased by the Noether's theorem of \eqref{sy:3} as
\begin{equation}\begin{aligned} 
\oint_\Omega dx \cdot  u \ \frac{d f (b)}{db} = \int d^3 x \frac{\partial f}{\partial (\partial_0 \phi^i)} \zeta^i_\Omega(\phi^j) \times \underbrace{\int_0^1 d\tau \frac{d\Omega^i}{d\tau} \delta^3\left(\phi^j -\Omega^j(\tau) \right)}_{\text{symmetry generator}},
\end{aligned}\end{equation}
where $ \Omega$ is the circulation path. To continue our analytical approach, we shall show how spin variables can be incorporated into the Lagrangian formalism before introducing spin Lagrangian. First, one argues that the energy density, pressure of the medium, and vorticity are not enough thermodynamical quantities to settle out a polarized system \cite{dm2,dm3}. Second, this new degree of freedom should act as a source breaking Kelvin's theorem. Following these lines, the last dynamical variable to define polarizable fluid is 
\begin{equation}\label{plv}
y_{\mu\nu} \sim u_\alpha \partial^\alpha \bigg( \sum_i \hat{T}^i \theta_i (\phi) \bigg) \equiv \chi ( b , \omega^{\mu\nu} \omega_{\mu\nu}) \omega^{\mu\nu},
\end{equation}
where $\hat{T}^i$ and $ \theta_i$ are the generators and local phase\footnote{The first interesting constrain is that solutions are no longer stationary as in the ideal Euler's equations. After a sufficient length of time, if the source is absent, the gradient will vanish, and thus the fluid will establish the homogeneous configuration.}, respectively, and the vortical susceptibility $ \chi(b,\omega^2)$ represents how inaccessible degrees of freedom coupled with macroscopic ones, turning out the fluid non-unitary by assumption because we cannot neglect feedback of microscopic variables. The relation between $y_{\mu\nu}$ and spin is the same as the relation between chemical potential and field phase in \cite{dm1}. Note that the nonexistence of Goldstone modes requires polarization parallel to vorticity in thermodynamical equilibrium \cite{dm1}.

Even though the formulation of hydrodynamics system via conservation of stress tensor leads to remarkable results near-equilibrium, the gradient expansion partially presents the fluid behavior since physics of fluctuation remains out of the partial differential equations \cite{kadanoff}. Moving on to a more "realistic" fluid, our effective action cannot be arbitrary, but must be in accord with symmetries ruling the problem under consideration. We begin with a complete description of the dissipative polarizable fluid by a generating functional
\begin{align}\label{gen:func:hydro}
Z[ b , y^2 ] = \int  \mathcal{D} b \mathcal{D} B_{ij}  \mathcal{D} y^{\mu\nu} \rho (\phi, y) e^{iS},
\end{align}
where the action $S$ describes the physical model proposed in our work 
\begin{align}
S &= S_{free} + S_{shear} + S_{bulk} + S_{pol} + S \left( \partial \phi_{\pm} \right)^2, \label{all:action} \\ 
% S_{free} &= \int d^4 x f (b) \\
S_{shear} &= \int d^4 x T^4_o   z_{IJK}(K^{\gamma} K_\gamma) b^2 B^{-1}_{ij} \partial^\mu \phi^{iI} \partial^\nu \phi^{jJ} \partial_\mu K^K_\nu {,}  \label{eqsshear} \\
S_{bulk} &= \int d^4 x T^4_o    h_{IJK}(K^{\gamma} K_\gamma) K^{\mu I} K^{ \nu J}  \partial_\mu K^K_\nu {,} \label{eqsbulk} \\
S_{pol} &= \int d^4 x T^4_o F (b(1 - c y^2 )).  \label{eqspol}
\end{align}
Remembering that the coefficients $\{ z_{IJK}, h_{IJK} \}$ becomes physical coefficients $\{ \bar{z}_{IJK}, \bar{h}_{IJK} \}$ if we remove the $CTP$ degeneracy $ \phi^{+K} = \phi^{-K}$ in the equations of motion. Following the guidelines of the variational principle, we assume all relations are valid locally. To see all considerations adopted in \cref{noneq} applied to hydrodynamics, we confine ourselves in the weak coupling limit, leading low-frequency, and long-wavelength order.

At the level of our effective field theory language, the dissipative construction of the shear viscosity $\eta$ requires additional variables. This transport coefficient breaks the volume-preserving diffeomorphism group in \cref{table:ss} (a uniform ball-shaped volume element has different dissipative forces than a stick-shaped inhomogeneous element of the same volume). It involves the linear introduction of the inverse matrix $B^{-1}_{IJ}$ which causes instability by turning the action unbounded from below. For bulk viscosity $\xi$, it is enough to double $K_\mu$ in \eqref{eqsbulk} since this dissipative mechanism still preserves the homogeneity, isotropy, and parity symmetry. Finally, to relate the local polarization phenomena, we perturb slightly the Lagrangian by introducing a symmetry breaking term in \eqref{eqspol}. This useful ingredient, forcing the internal symmetry from $SO(3)$ to $SO(2)$, arises a well-defined polarization with intrinsic anisotropic degrees of freedom, see \cref{table:ss}. With the help of the energy positivity $y^{\mu\nu} y_{\mu\nu} > 0$ and $ y^{\mu\nu} u_{\mu}= 0 $, we have to restrict the polarization form. Furthermore, the physical origin of this "new" induced polarization comes from linking vortex and spin \eqref{plv}. The convenient way, from the first principle, to evaluate this transport property is by assuming an external time-independent vortex-source. This forced symmetry breaking reduces each cell fluid to $SO(2)$ symmetry and introduces a degeneracy that changes the structure stability. We will postpone the stability question for \cref{pmc} when we will include relaxation terms.

\begin{table}[H] 
\centering
\begin{tabular}{c|ccccc}
&parity & time & charge & $SO(3)$ & $S\text{Diff}\,(\mathbb{R}^{1,3})$\\
\hline 
Perfect fluid & even & even & even & unbroken & unbroken  \\
Shear viscosity & even & even & even & unbroken &  broken  \\
Polarization & odd & odd & even & broken  & unbroken  \\ 
\end{tabular}
\caption{Symmetries of various terms beyond local equilibrium hydrodynamics.}
\label{table:ss}
\end{table}

\section{Linear response theory for dissipative spin hydrodynamics}\label{chap:3}

\subsection{Fluctuation-dissipation theorem}\label{flueft}

Effective theory is an expansion build around a clear separation of energy scales. We can develop it by writing down all terms allowed by the unbroken symmetries of the "ultraviolet" theory, and weighting each term with a factor encoding its length scale (which could be the infrared over the ultraviolet scale, or intermediate-scales arising from condensates, the latter are particularly characteristic of broken symmetries). Hydrodynamics admits three possible length scales, in a hierarchy
\begin{equation}\label{hyerarchy}
\underbrace{l_{micro}}_{\sim s^{-1/3}} \ll \underbrace{l_{mfp}}_{\sim \eta/(s T_0)} \ll \underbrace{ l_{hydro}}_{\sim (\partial u)^{-1}},
\end{equation}
where $l_{micro}$ encodes microscopic correlations, parameterizing the lack of molecular chaos and the presence of entanglement. It scales due to distance of separable microscopic degrees of freedom, which depends on the cube root of the entropy density $s$. The mean free path $l_{mfp}$, parameterizing macroscopic dissipation, and related to the sound wave dissipation length, depending on viscosity, entropy density, and microscopic temperature $T_0$.

To access the complex dynamics structure, which lies in a non-equilibrium system, we shall delimit the energy scale that our theoretical framework can track relevant variables. We stress out the number of inaccessible degrees of freedom increase as long as the system accuracy. Hence the parameters responsible for relevant interactions control the power-counting rule of effective field theories. So we require a well-defined spatial and time separation as in \eqref{hyerarchy}, yet what we are interested in is the ratio between these three length scales. 

From the effective field theory viewpoint, the dynamical interpretation of dissipation is provided by the flux across the cutoff $\Lambda$ in momentum space. Under such circumstances, the ideal fluid is a unitary system that describes the variation of hydro-variables $\phi$ with slow dynamics $\partial_\mu \sim l^{-1}_{hydro}$ because the correlation of microscopic interactions remains neglecting below $\Lambda$. Even if the dissipative fluid dynamics of first ($NS$) and second order ($IS$) in Knudsen number $\textbf{Kn} \equiv l_{mfp} / l_{hydro} $ emerge from a system close to the equilibrium, the excitations of spin cannot still influence the fluid parametrization. We then consider these conditions a clear separation between micro- and macro-length for what demanding the ratio $l_{micro}/l_{hydro} $ be neglected. The usual approach that effective Lagrangian \eqref{mix::ctp} could recognize spin as a viable candidate is by the symmetry and typically length scale \cite{dm1}. In local fluid cells, the vortices $\mathcal{O}(l^{-1}_{vortex})$ are tightly coupled with spin inside local fluid cell \cite{lisa}. Beyond that, when we request causality and stability conditions, vortex dimensions has no minimal arbitrary size, which means a minimal dissipation or length to vortex-coupling\cite{dm3}\footnote{This fate for our spin hydrodynamics characterizes $l_{vortex} \ll l_{hydro}$ even if it reaches the thermodynamic equilibrium.}. Such lower limit request a new power-counting scheme, see discussion in \cite{foca}. However, hydrodynamics of fluids with spin works well when microscopic collisions are no longer valid to represent the spin transport process\cite{Becattini2018duy}. Using these constraints and the spin length fluctuation is $\ll l_{mpf}$, we may assume without lost of generality $l_{vortex} \sim l_{mfp} $. Moreover, because of $\chi(b,\omega^2)$ in \eqref{plv}, the transference between angular momentum and spin demands 
\begin{equation} 
\frac{l_{micro}}{l_{mfp}} \sim \frac{l_{mfp}}{l_{hydro}} \equiv \textbf{Kn}. 
\end{equation}
Even though the physics establishes distinct ultraviolet (microscopic) and infrared (macroscopic) scales \eqref{hyerarchy}, the $\textbf{Kn}$ allows us to treat in the same foot the transverse momentum ($\sim l_{mfp}/l_{hydro}$) and vortex-spin coupling ($\sim l_{micro}/l_{mfp}$).

As we advocated in this article, the polarizable hydrodynamics models \cite{palermo,bec,flork1,shearspin,gale,hongo} dismiss the Navier-stokes ($NS$) viscosity in the phenomenological equations of motion. The motivations for introducing such processes rely on theoretical and experimental perspectives. One of them is to provide a more realistic role for predicting the spatial distribution of spin \cite{chen2019qzx}, to correct the analytical predictions of polarization momentum dependence \cite{Adam}, to claim dissipative effects of $RHIC$ fluid \cite{Heinz}, and for outstanding theoretical properties. We proceed with the approach introduced in \cref{noneq}, which includes angular momentum as per the prescription of \cite{palermo}, to the density matrix  
\begin{equation}
\rho =\mathcal{Z}^{-1} \exp{[- \beta( \mathcal{E} - v \cdot p - \frac{1}{2} \varpi_{\lambda \nu} \cdot J^{\lambda \nu } )]}
\end{equation} 
where the partition function $\mathcal{Z}$ is the normalizing constant, $v$ is spatial velocity, $p$ is momentum density, $\mathcal{E}$ is     Hamiltonian,  $\beta = 1/T$ is inverse of temperature,  $\varpi_{\lambda \nu}$ is vorticity, and $J^{\lambda \nu }$ is finite total angular momentum. The $\rho$, to be that of a thermal bath, denotes an ensemble of states with $Tr[\rho] = 1$. By disturbing the energy density, we have 
\begin{equation}\label{old:energy}
\delta \mathcal{E} (t) = \int d^3x \bigg[ \frac{\delta T}{T} (t, \textbf{x}) \epsilon (t, \textbf{x}) + p^i  (t, \textbf{x}) v_i  (t, \textbf{x}) + Y^{\mu\nu} (t, \textbf{x})  \omega_{\mu\nu} (t, \textbf{x} ) \bigg].
\end{equation}
where $\epsilon$ is the energy density. We discard terms up to second order because the thermodynamics forces are small within the limit of the near-equilibrium state. Any operator $\mathcal{A}$ evolves in according with 
\begin{equation}\label{take}
\rho (t) = U (t) \rho_o (t) U^{-1} (t),
\end{equation}
where $U(t)$ is a unitary matrix. After taking an average over the equilibrium ensemble, becomes 
\begin{equation}
\langle \mathcal{A} \rangle = Tr [ \rho \mathcal{A} ] =  \langle \rho_{eq} \ U^\dagger (t,t_0) \  \mathcal{A} \ U(t,t_0)  \rangle,
\end{equation}
where $\rho_{eq}$ is the equilibrium density operator. The hydrodynamic perturbation can be introduced using the interaction picture $U(t,0) = \mathcal{T} \{ \exp \left(-i \int^t_{0} \mathcal{E} (t') dt' \right )  \}$ to obtain the perturbative information of how deviations out of equilibrium disappear. Following the linear response approach, the expectation value is calculated via 
\begin{equation}\label{old:green}
\delta \langle \mathcal{A} (t, \textbf{x}) \rangle = \langle \mathcal{A} (t, \textbf{x}) \rangle -  \langle \mathcal{A} (t, \textbf{x}) \rangle_{eq} =  \int d^3 \textbf{x}' \int^t_{- \infty} d t'  e^{\alpha t} \Theta (- t') \langle \mathcal{A} (t, \textbf{x}) , \delta \mathcal{E} (t',\textbf{x}^\prime) \rangle_{eq.},
\end{equation}
where $ \langle \mathcal{A} \rangle_{eq} = Tr [\rho_{eq} \ \mathcal{A}]$ and  $\alpha$ encodes the adiabatic switching operation for the external source. The fluctuation-dissipation theorem establishes the response to small perturbation as the correlation between the perturbed observable and another conjugated one concerning to \textit{energy} in \eqref{old:energy}. The retarded and advanced Green's functions are defined via the Heaviside function 
\begin{equation}
G_{R/A} (t-t',\textbf{x} - \textbf{x}') = -i \Theta\left(\pm (t - t')\right) \langle[ \mathcal{A} (t, \textbf{x} ), \mathcal{A} (t' , \textbf{x}' ) ]\rangle.  
\end{equation}
Beyond that, we shall discuss the general concept of dissipative current around local equilibrium from \eqref{old:green}
\begin{equation}\label{old:GK}
J_\alpha (t)=\int d^4 x' G_{\alpha \beta } (t-t') F_\beta (t')  + \mathcal{O}(F^2),
\end{equation}   
where the Green's function $G_{\alpha \beta} (x-x')$ depends on local currents $J_\alpha (x)$ and external thermodynamical sources $F_\beta (x)$. Note the currents corresponding to $ \{ Y^{\mu\nu} , p^i_T , p^i_L \} $ have as the sources $ \{ \omega^{\mu\nu}, v^i_T ,v^i_L  \} $. The equation above helps expressing the linear relations of dissipative currents in terms of thermodynamics forces asymptotically close to equilibrium. Nonetheless, these same linear equations rule, in the statistical equilibrium, the fluctuations decaying of the steady-state. Thus we cannot predict if a perturbation is created by a fluctuation or a small external thermodynamical source. This observation elucidates that different classes of independent elementary mechanisms involved in the irreversible integral \eqref{old:GK} yield the same expected result as a consequence of microscopic reversibility. We then categorize this property according to the famous Onsager reciprocity $G_{\alpha \beta} = G_{\beta \alpha}$, benefited of the detailed balance \cite{kadanoff}.

\begin{figure}[H]
\centering 
% of the shear stress tensor,
\includegraphics[width=15cm]{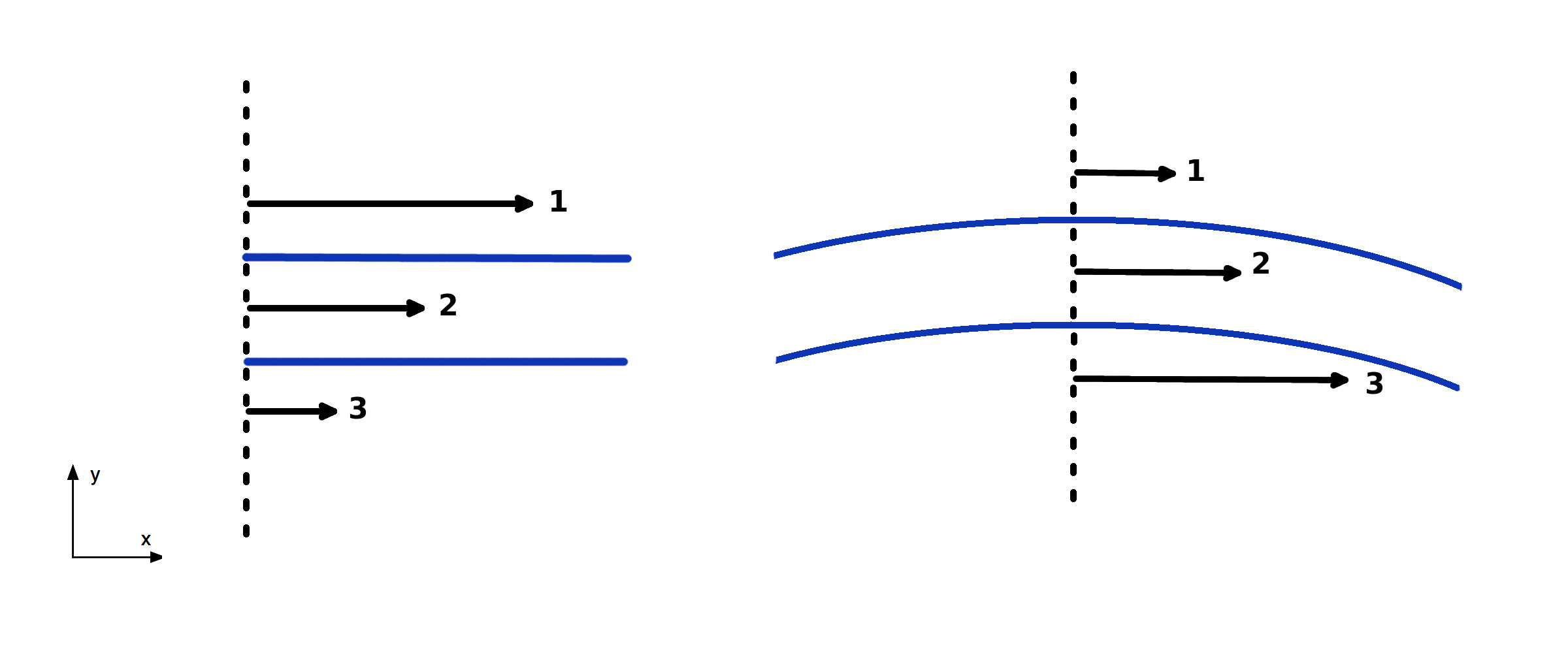}
\caption{A schematic view describing the velocity profile shape, arrows, for parallel (left) and rotating (right) viscous fluid. The former follows  $\sigma_{xy} \sim \eta \frac{\partial v_x}{\partial y}$, and $y$ is perpendicular to the interface, while the latter adopts the azimuthal $\psi$ direction  $ \sigma_{xy} \sim  \sigma_{r \psi} \sim ( \frac{ \partial v_\psi}{\partial r} - \frac{v_\psi}{r}) = \eta r \frac{\partial \Omega}{\partial r}$, and $r$ is the radius.}
\label{viscfig}
\end{figure}

The foregoing discussion allows us to elucidate new physical phenomena from the interaction of polarization, shear, and bulk viscosity. Following the macroscopic perspective, let us begin with the average polarization  
\begin{equation}\begin{aligned}\label{old:pol} 
& \langle Y^{\mu\nu}  (t, \textbf{x}) \rangle_{\omega} -  \langle Y^{\mu\nu}(t, \textbf{x})   \rangle_{eq,\omega=0}  \approx \ + \ i \int dt^\prime  \int_V d^3 x^\prime \langle [Y^{\mu\nu}(t,\textbf{x}),Y^{\alpha\beta}(t^\prime,\textbf{x}^\prime)] \rangle_{eq}    \omega_{\alpha\beta} \  + \\
& i \int dt' \int_V d^3 x \langle [Y^{\mu\nu}(t,\textbf{x}), p^i_T (t^\prime,\textbf{x}^\prime)] \rangle_{eq} v^i_T \ 
+ \ i \int dt' \int_V d^3 x \langle [Y^{\mu\nu}(t,\textbf{x}), p^i_L (t^\prime,\textbf{x}^\prime)] \rangle_{eq} v^i_L,
\end{aligned}\end{equation}
where the $p^i_T$ and $p^i_L$ are the transverse and longitudinal momentum flux, respectively. The last two terms represent the polarization coupling with transverse and longitudinal momentum, respectively. The former is the shear-induced by polarization \cite{shearspin}, while the latter is the anisotropic expansion from the fluid. Before dealing with these cases, we shall recall the viscosity gradient forces tend to disappear the velocity deviations from fluid layers. This friction effect, a mechanism of momentum transfer between adjacent fluid layers, produces distinct velocity distribution curves, as shown qualitatively in \cref{viscfig}. For rotational flows, it expresses the velocity increases from outer annuli layers to inner ones, and so the dissipation rate depends on the height disc. We can assume the shear viscosity plays a role in the spatial structure of the spin by redistributing its location in the fluid. In practice, the spin current generated by the vortex fluid is deformed because the spin loses angular momentum to the environment (fluid) and moves inward, whereas the shear viscosity transfer angular momentum to outer layers.

The computation of the irreversible observable \eqref{old:pol}, by means of a statistic correlation function, can gather varieties of dissipative phenomena. Nonetheless, these three natural fluctuations will not have the same weight in all physical problems as a result of considering both initial and boundary conditions, as well as the free energy scale. Moreover, the domain of small viscosity and weak coupling, in linearized theory, restrain the interaction among the fluctuations above, only proposed for the non-equilibrium steady state in non-linear circumstances. As we restrict ourselves to the first-order processes near-thermal equilibrium, the cumulative effects, due to different external sources, also run over spin in reverse trajectory (microscopic reversibility). Hence, we derive the average transverse momentum density as
\begin{equation}\begin{aligned}\label{for:2}
& \langle p^i_T (t, \textbf{x}) \rangle_{\omega} - \langle p^i_T (t, \textbf{x})  \rangle_{eq,\omega=0} \approx   \\
&  i \int dt' \int_V d^3 x \langle[ p^i_T (t,\textbf{x}), p^j_T (t^\prime,\textbf{x}^\prime)]  \rangle_{eq} v^j_T + i \int dt^\prime  \int_V d^3 x^\prime \langle [p^i_T (t,\textbf{x}), Y^{\alpha\beta} (t^\prime,\textbf{x}^\prime)] \rangle_{eq} \omega_{\alpha\beta}.   
\end{aligned}\end{equation}
The last term, which exhibits a vorticity resistance, is called the rotational shear viscosity, a mechanism where the vortex losses angular momentum due to the friction of rotational fluid. As the spin lay out a vortex structure in the rotational fluid, the vortices interfere with the propagation and scattering of sound waves. The observation of this phenomenon was already studied in the literature \cite{liter}. Last but not least, the average longitudinal momentum density is
\begin{equation}\begin{aligned} 
 & \langle p^i_L (t, \textbf{x})  \rangle_{\omega} - \langle p^i_L  (t, \textbf{x}) \rangle_{eq,\omega=0} \approx  \\ 
&  i \int dt' \int_V d^3 x \langle [ p^i_L (t,\textbf{x}),  p^j_L (t^\prime,\textbf{x}^\prime)] \rangle_{eq} v^j_L + i \int dt^\prime  \int_V d^3 x^\prime \langle [p^i_L (t,\textbf{x}), Y^{\alpha\beta} (t^\prime,\textbf{x}^\prime)] \rangle_{eq}  \omega_{\alpha\beta},   
\end{aligned}\end{equation}
where the first correlation function measures the deviation from equilibrium pressure, while the other one determines the anti-symmetric pressure from spin compression.

In principle, to achieve a complete description of the phenomenological relations, we shall establish further assumptions, more precisely at first-order, in accordance with Markovian diffusion. For that reason, the average of any operator at the equilibrium stage must satisfy defined thermodynamic relations in \eqref{old:pol} at $t < 0$ as $ \partial \ln{\mathcal{Z}} / \partial \vec{\omega} = \langle \boldsymbol \vec{J} \cdot \boldsymbol \hat{\omega} \rangle /  T$ and $ \partial \ln{\mathcal{Z}} / \partial v^j = \langle p \cdot \hat{v} \rangle \hat{v}^j / T$. Hence these thermodynamic derivatives in the long wavelength limit are  $\lim_{\textbf{k} \to 0} \chi  = \partial Y^{\mu\nu} / \partial \omega^{\mu\nu}$ and $\lim_{\textbf{k} \to 0} w_o  = \partial p^i / \partial v^i$, where $w_o$ is enthalpy, with the initial conditions expressed appropriately as
\begin{equation}
p^i_{T,L} (0, \textbf{x}) = w_o u^i_{T,L}, \ \  \partial_t \langle p^i_{T,L} (0,\textbf{x}) \rangle_{|t=0} = 0, \quad \langle Y^{\mu\nu} (0,\textbf{x}) \rangle = \chi  \omega^{\mu\nu}, \quad  \partial_t \langle Y^{\mu\nu} (0,\textbf{x}) \rangle_{|t=0} = 0.   
\end{equation}

\subsection{Variational approach}
In this subsection, we present an alternative tool to interplay a linear relation between dissipative hydrodynamics and spin. The qualitative aspects from the previous method \cref{flueft} are insufficient to provide a physical meaningful for generic two-point correlation functions. Even though thermodynamical identity and conservation laws are the tools for writing fluctuation-dissipation theorem, previous studies already anticipated the observable as spin requires knowledge which goes beyond the phenomenological equations \cite{shearspin}. For instance, the correlation of acceleration spin $Y^{0i}$ and energy density $\epsilon$ cannot be determined from the perturbation of an external force on thermodynamical identity. To bypass this difficulty, we suggest another approach in which metric fluctuations rather than external sources originate the small deviation from equilibrium. We introduce in \eqref{old:green} the stress tensor operator 
\begin{equation}\label{var:prin}
\delta \langle \mathcal{T}^{\mu\nu} (t,k_i) \rangle =  \int^{\infty}_{- \infty} d t'  \Theta (-t') e^{\alpha t'} G^{\mu\nu}_{\lambda\delta}(t-t',k_i) \mathcal{S}^{\lambda\delta}(t',k_i), 
\end{equation}
where the new source $\mathcal{S}^{\lambda\delta}$ encodes not only metric fluctuation $h^{\mu\nu}$ but also gauge field $\omega^\mu$. We can replace the energy-momentum tensor for the conserved current $\mathcal{J}^{\mu\nu}$. We explicitly rewrite \eqref{var:prin} in our argument by using the variational principle \cite{Kovtun2012rj}
\begin{equation}\begin{aligned}\label{eq:GR-sources}
G^{\,R}_{T^{\sigma\tau} T^{\mu\nu}}  & = - 2 \left.\frac{\delta {\cal T}^{\sigma\tau}  }{\delta h_{\mu\nu} }\right|_{h^{\alpha \beta} = \omega^\alpha=0}, \quad \ \  G^{\, R}_{J^\mu \! J^\nu}   = - 2
\left.\frac{\delta {\cal J}^{\mu}  }{\delta \omega_{\nu} }\right|_{h^{\alpha \beta}=\omega^\alpha=0}, \\ 
G^{\,R}_{T^{\sigma\tau} T^{\mu\nu}}  & = - 2 \left.\frac{\delta {\cal T}^{\sigma\tau}  }{\delta h_{\mu\nu} }\right|_{h^{\alpha \beta} \equiv \omega^\alpha=0}, \quad G^{\,R}_{J^\sigma T^{\mu\nu}}= -2
\left.\frac{\delta {\cal J}^{\sigma}  }{\delta h_{\mu\nu} }\right|_{h^{\alpha \beta}=\omega^\alpha=0}.
\end{aligned}
\end{equation}
One can see that, by inspection, these equations give the equivalent of the propagator in \eqref{propagator1} in the first order. Such definition provides the functional differentials are well-defined 
\begin{subequations}\begin{align} 
\omega (G_{T^{x0},T^{x0}} (\omega,k_x) + \underbrace{ f_y y - f}_{\epsilon} ) &= k_x G_{T^{x0},T^{xy}} (\omega,k_x),  \\
\label{eq22}
\omega (G_{T^{x0},T^{xy}} (\omega,k_x)) &= k_x (G_{T^{xy},T^{xy}} (\omega,k_x) + \underbrace{f - f_b b}_{P}),  
\end{align}\end{subequations}
where $P$ is the fluid pressure. Note that the contact terms appear naturally because of the reminiscent "functions" $\delta(\omega)\delta^3 (\textbf{k})$. Let us then sum the two equations above
\begin{equation}
\omega^2 G_{T^{x0},T^{x0}}(\omega,k_x) - k_x^2 G_{T^{xy},T^{xy}} (\omega,k_x) = 0.
\end{equation}
Here, by suppressing the contact terms, we satisfy the conservation law. In long-wavelength, the correlation function decays to the asymptotic configuration where the well-known thermodynamic quantity emerges  
\begin{equation}
\lim_{k_x \rightarrow 0 } G_{T^{x0},T^{x0}} (0,\textbf{k})  = \int d^3 \textbf{x} \int^{\beta}_0 d t \ \langle T^{x0} ( t, \textbf{x} ) T^{x0} (0) \rangle  = \beta ( \langle  \mathcal{E}  T^{x0} \rangle - \langle  \mathcal{E} \rangle \langle  T^{x0} \rangle).  
\end{equation}
The enthalpy can be explicitly written as
\begin{fleqn}
\begin{equation}\label{true0}
\quad \ \lim_{\omega \to 0} \lim_{k \to 0} \frac{1}{\omega} Re G^{R}_{T^{0x},T^{0x}} (\omega,\textbf{k}) = (\epsilon + P ) = f_b.
\end{equation}
\end{fleqn}
The explicit relation between the retarded Green function and the transport coefficient originated from Kubo formulae are 
% \begin{fleqn}
\begin{eqnarray}\label{true1}
&& \frac{1}{12} \lim_{\omega \to 0} \lim_{k \to 0}  \partial^3_\omega Re G^{R}_{T^{0x},T^{0x}} (\omega,\textbf{k}) = \chi^2, \\  
&& \lim_{\omega \to 0} \frac{1}{\omega} \lim_{k \to 0} Im G^{R}_{T^{0x},T^{0x}} (\omega,\textbf{k}) = f_b + b_o^{3} (2 \Bar{z}_{003} - 3 \Bar{z}_{033} + 4 \Bar{z}_{303} + 3 \Bar{z}_{333}), \\ \label{true3}
&& \lim_{\omega \to 0} \lim_{k \to 0} \frac{1}{k} \partial_\omega G^{R}_{J^x,\omega^{xy}}   =    b_o^3  (\bar{h}_{003,03} +\bar{h}_{303,03}+\bar{h}_{333,03} + \bar{h}_{033,03} +2 \bar{h}_{300,03}) - 4 b^5_o (\bar{h}_{333,00} + \bar{h}_{303,00})   
\nonumber \\
&& + \ 4 b^5_o (\bar{h}_{003,33} + \bar{h}_{303,33} + \bar{h}_{333,33} + \bar{h}_{033,33} + \bar{h}_{330,33} + \bar{h}_{300,33}).
\end{eqnarray} 
% \end{fleqn}
The \crefrange{true1}{true3} are valid within a limited range (near-equilibrium) and fail to include non-perturbative analysis. We have then identified the relevant couplings from rotational dissipative fluid in terms of the effective action expansion \eqref{all:action}.

\subsection{Interactions of hydrodynamic modes}

In the previous section, the results obtained are governed by a linearized hydrodynamic regime. They concern an asymptotic region where slow internal fluctuations of hydrodynamical variables are characterized by an averaging over the equilibrium ensemble. To extract more dynamical information, we can analyze the correlation function behavior in more complex circumstances. In a sense, the structures closely connected with thermal fluctuations are a dominant factor in studying hydrodynamics beyond first-order. Hence, with the help of non-hydrodynamics fluctuations, one can investigate the collective modes appearing in a limited region $(\omega, \textbf{k})$, sensitive to microscopic dynamics. Furthermore, one can show that for second-order deviation from the equilibrium, the hydrodynamics modes are subjected to non-analytical conditions where the time correlation function decays at long-time tails.

The main point is to understand how vorticity effects and spin thermal fluctuations interplay in the dissipative polarizable fluid. For now, we continue to look at the system from the macroscopic perspective and follow the approach of \cite{kovtail}, adapted to the Lagrangian picture. Examining the stress tensor in the co-moving frame, up to second power of hydrodynamical variable, we have 
\begin{equation}\begin{aligned}\label{tensordef}
T^{pq} \sim & \, ( c_s^2 [\partial \pi^2 ] + \frac{1}{2}(1 + c_s^2)( \dot \pi [\partial \pi] + \dot \pi^2 ) +  \frac{1}{6}(3 c_s^2 + f_3 ) [\partial \pi]^2 ) \delta^{pq} - 2 (z_{00K} + h_{00K,00}) \delta^p_k [\partial \dot \pi] \partial^q \pi^{kK}  \\ 
& + z_{IJ0} (\pi^{mI} \pi^{nJ} S^{pq}_{mn} + ( \dot \pi^2  + \frac{1}{2} [\partial \pi]^2 - [\partial \pi \cdot \partial \pi] ) \delta^{pq} \delta^{I}_J + (\dot \pi \cdot \partial \pi^{pI} - [\partial \pi] \dot \pi^{pI}) \delta^{qJ} ) + 2 \, F_b   \chi^2 \times   \,   \\
&   (  g_{i[\rho} \, \partial_{0]} \, \dot \pi^i \ddot \pi^\rho \delta^{pq}  + \, \dot \pi^i H^{pq}_{ij} \, \dot \pi^j )  +  2 F_b    \chi \ y^p_{\rho} \delta^{0q} + \frac{c^2_P}{3} F_b   y^p_\sigma \epsilon^{q \rho \alpha}\epsilon_{0 \rho i} \partial^\sigma \partial_\alpha \pi^i + \dots, 
\end{aligned}\end{equation}
where $f_3$ corresponds $f^{\prime\prime\prime}$, $c_s$ is the sound speed, $c_P$ is the longitudinal perturbation emitted by vortex-spin source, and the projectors read $S^{ab}_{mn} = \frac{1}{2}(\delta^a_m \delta^b_n + \delta^a_n \delta^b_m - \frac{2}{3} \delta^{ab} \delta_{mn})$ and  $H^{ij}_{kl} = ( \delta^i_k \delta^j_l - \delta^i_l \delta^j_k )$. The spin current is
\begin{equation}\begin{aligned} 
J^p \, \sim & \, ( \dot \pi^2 \delta^p_0 - \frac{1}{2}( c_s^2 - f_3) \partial^p \pi \cdot \dot \pi ) + \frac{1}{6} ( h_{0JK,00} + z_{0JK} )  \dot\pi^{J} \cdot ( \dot \pi^{K} \delta^p_0 + \partial^p \pi^{K }) + \frac{1}{6} z_{IJ0} \epsilon^{pab} \epsilon_{0mn} \partial_a  \pi^{mI}   \\ 
& \times \partial_b \pi^{nJ} + 2 z_{IJK} ( \dot \pi^{I} \cdot \dot \pi^{J} \delta^{pK} +  \dot \pi^I \cdot \partial \pi^{pJ}  \delta^{K}_0 ) + z_{000} ( \dot \pi^2 + [\partial \pi ]^2 - \partial^p [\partial \pi^2])  + w_o c_p^2 F_b \chi \partial^p \partial_j \pi^j     \\ 
& -  2 w_o F_b  ( \chi^2  \partial^p \dot \pi^I \cdot \partial \dot \pi^I  - \,  \chi \, \partial^p \chi [ \partial \dot \pi \cdot \partial \dot \pi ] \,  )  + \dots. \\ 
\end{aligned}\end{equation}
This current, a conserved dynamical variable, presents the irreversible flux as a decomposition of $y$-polarization $J_P$ and $\pi$-hydro $J$ currents. The dots contain third or higher-order terms of $\pi$. The physical meaning of $T^{pq}$ and $J^p$ is to explore the influence of collective excitations on transport coefficient values. These short-lived modes involve non-equilibrium states at an early time, which corrects the bare Green's function in \crefrange{true1}{true3}. Following the Onsager's reciprocal principle, the fluctuation-dissipation theorem lies in regions where the stable steady states have a well-distinct small and large-scale time \cite{kadanoff}. Consequently, to restore the finite result on the gradient expansion scope, we assume Gaussian fluctuations (Markovian dynamics) to evaluate the quadratic-order correlation from \eqref{tensordef}
\begin{eqnarray} \label{G:2}
G^{(2)}_{T^{ij}T^{kl}} (t,\textbf{x}) &&= \frac{T}{ w^2_0 } \langle \pi_i (t,\textbf{x}) \pi_j (t,\textbf{x}) \pi_k (0) \pi_l (0) \rangle_{eq}, \nonumber \\ 
&&= \frac{T}{ w^2_0 } \, \langle \pi_i (t,\textbf{x}) \pi_k (0) \rangle_{eq} \langle \pi_j (t,\textbf{x})  \pi_l (0) \rangle_{eq} , \nonumber \\
&&= \frac{2 T}{w_o^2} \int \frac{d \omega^\prime}{2 \pi} \int \frac{d^3 k^\prime}{(2 \pi)^3} G^{(0)}_{ T^{0i} T^{0j} } (\omega^\prime , \textbf{k}^\prime ) G^{(0)}_{ \, T^{0k} \, T^{0l} } (\omega - \omega^\prime , \textbf{k} - \textbf{k}^\prime),
\end{eqnarray} 
factorizing the Green function as the product of two zero-order ones. In particular, we evaluate \eqref{G:2} in the long-wavelength limit $ \chi^2 \textbf{k}^2 \ll \gamma (z,h) \textbf{k} \ll c_s$ by 
\begin{equation}\begin{aligned}
G^{(0)}_{ \, T^{0i} \, T^{0j} }(\omega, \textbf{k}) = & \ \frac{w_o T}{2}  \bigg[  \bigg( \delta^{ij} - \frac{k^i k^j}{\textbf{k}^2} \bigg) \frac{  \textbf{k}^2 - \gamma_\eta \omega^2 \textbf{k}  - 2  \chi^2 (\omega^4 - \textbf{k}^2 \omega^2) }{f_b \omega^2 - i \gamma (z,h) \omega \textbf{k}^2 - 2 \chi^2 \omega^4  } \ + \\   &  \bigg(\frac{k^i k^j}{\textbf{k}^2} \bigg) \frac{ \omega^2 + \gamma_\eta \omega \textbf{k}^2  - \chi^2 (3 \omega^4 + 2 \omega^2  \textbf{k}^2 )}{f_b (\omega^2 - c_s^2 \textbf{k}^2 ) - i \gamma_\eta (2 \omega^2 \textbf{k}  + 3 \omega \textbf{k}^2 ) - i \gamma (z,h) \omega \textbf{k}^2 - 2 \chi^2 (\omega^4 - \textbf{k}^2 \omega^2)}  \bigg],
\end{aligned}\end{equation} 
where $\gamma (z,h) =\gamma_\eta (z) + \gamma_\xi (h)$ is the function composed by elementary components of the shear $\{z_{IJK}\}$ and bulk $\{h_{ijk,lm}\} $ viscosity, respectively. Making the following association with \eqref{G:2}, we obtain  
\begin{equation}\label{billy}
\mathcal{V} \langle \{ T^{ij} (t,\textbf{x}) , T^{kl} (0) \} \rangle =  \frac{T}{w_o^2}
(S^{ij}_{mn} S^{kl}_{pq} + H^{ij}_{mn} H^{kl}_{pq}) G^{(2)}_{T_{mn} T_{pq}} (t,\textbf{x}),
\end{equation}
where $\mathcal{V}$ is the spatial volume. For $t>0$, taking the limit at zero momentum and low frequency to characterize the hydrodynamic regime, we obtain 
\begin{equation}\begin{aligned}\label{old:non}
\mathcal{V} \langle \{ T^{ij}  (t,\textbf{x}) , T^{ kl} (0) \} \rangle \sim \ & \mathcal{O} (\Lambda) + H^{ij}_{kl} \frac{T^2}{6 \pi} \bigg(   \frac{\chi^{-5/2}}{w_o t^{3/2}  } +  \bigg[ 3 + \bigg( \frac{1}{6} \bigg)^{1/2} \bigg]  \frac{\chi^{-3/2}}{2 t^{1/2}  } \bigg) - \frac{S^{ij}_{kl} }{60 \pi} \bigg[ 7 + \bigg( \frac{3}{2} \bigg)^{3/2} \bigg] \times  \\
& \  \frac{1}{ (\gamma_\eta (z) t)^{3/2}} + \text{(exponential decay)}.
\end{aligned}\end{equation}
Before starting this discussion, let us analyze the symmetric properties of the matter. The broken rotation group in \cref{table:ss} determines the spin alignment (orientation) as a new hydrodynamic degree of freedom responsible for describing the equilibrium thermodynamics state. This physical quantity acts as a non-conservative source generating a non-vanishing current from the angular momentum conversion into spin momentum. We found this current-current correlation, more precisely the auto-correlation of spin velocity $\langle v_s (t) , v_s (0) \rangle$, decays by a non-analytical time power-law $\sim t^{-3/2} $. Supposing at $t=0$, the average velocity of a spin particle $ \langle v_s \rangle^2$ decreases slowly through an elliptical shell region in the neighborhood of a rotating fluid. Since the randomization motion governs this whole process, the mechanism of vorticity diffusion is responsible for spreading out the average spin velocity in the elliptical shell column. As the neighborhood rearranges at a time $\tau$ longer than the spin takes to approach a "local equilibrium" inside this column of rotational fluid. We expect, therefore, the decaying of $ \langle v_s \rangle^2$ be $ t \sim A /(\theta \times r)^2 \tau $, with the azimuthal angle $\theta$, and the transverse area of elliptical shell column $A$. Recalling this diffusion process, shear momentum induced from vortical susceptibility depends on the shell surface radius \cref{viscfig}   
% inside of rotational fluid

The $\Lambda$ represents the correction for transport coefficient from shear and sound collective modes. Clearly, we label the two exponential contributions as the fast variable in which the first one is connected with the relaxing mode of shear dissipation, while the other one is related to vorticity diffusivity. The contribution of each separable sum represents a conserved quantity of collective modes, which cannot be neglected for $l_{micro}$ scale. Indeed, we can identify these modes as emerging from length scales $\sim$ microscopic degrees of freedoms, not-generated by old-fashioned hydrodynamic \cite{Landau}. This relevant aspect can be understood as the breaking down of gradient expansion signature for out-of-equilibrium systems. Since the macroscopic distance and time determine that collective behavior can be expressed as a continuum medium, the correlation function of \eqref{billy} arises from features beyond the hydrodynamic regime. In this case, the fluctuation \eqref{old:non}, at initial condition, is sensible to the presence of non-analytical and exponential terms. The interest in such phenomena is notable in various branches of physics as the perception of vorticity in the hyperon polarization measurement (STAR experiment) \cite{34}, condensed matter \cite{35}, and higher energy \cite{Heller}.

\section{Feynman diagrams for polarizable fluid with dissipation}\label{chap:4}

We apply the ideas developed in \cite{uslag} and \cite{DM} to explore the behavior of a dissipative polarizable fluid. Even though our effective action \eqref{all:action} in $(3+1)$ dimensions is non-renormalizable because of the coupling $\chi$ with negative mass dimension, it should not be interpreted as ultraviolet completion of a spinless fluid. This section turns out to be a relatively straightforward matter for adding shear and spin Lagrangian together. This procedure serves as a theoretical guide to investigate the central idea of how the polarization current modifies with the inclusion of well-know dissipative effects. In fact, the Feynman diagram is a powerful tool that keeps track of classical phenomena at tree-level amplitudes. Our independent DoFs are summarized in the table below

\begin{table}[htbp]
\centering
\begin{tabular}{l|ll}   
  & Feynman propagator &  \\
\hline Transverse excitation & \parbox[c]{1em}{  \includegraphics[width=1in,height=1em]{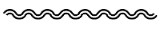}} \\
  Longitudinal excitation  & \parbox[c]{1em}{
  \includegraphics[width=1in]{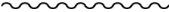}} \\
  Transverse Polarization  & \parbox[c]{1em}{
  \includegraphics[width=1in]{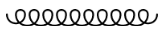}} \\
  Longitudinal Polarization  & \parbox[c]{1em}{
  \includegraphics[width=1in]{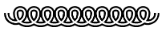}} \\
\end{tabular}
\caption{We list each Feynman line of polarizable fluid with dissipation (See \cref{twop} for more details).}
\label{tab:stimuli}
\end{table}

%For completeness, EFT techniques are more practical to study fundamental properties of fluid and derive a consistent theory.

\subsection{The linearized effective action}

Let us incorporate perturbative effects by expanding at the long-wavelength and low frequencies, the Lagrangian \eqref{all:action} up to fourth-order. We can easily separate the perturbative information of free hydro-modes 

\begin{fleqn}
\begin{eqnarray}\label{fdsa}
\quad \ \ \mathcal{L}_{\text{free}}  &=& \ w_o f_b \bigg( \frac{1}{2} \dot {\pi}^2 - \frac{1}{2} c_s^2 [\partial \pi]^2  + \frac{1}{2} [\partial \pi^T \partial \pi ]) - z_{0JK}   (   \dot \pi^J \cdot \partial [\partial \pi^K ] +   \partial^2 \pi^J \cdot \partial \pi^K  + [ \partial \pi^J \cdot  \partial \dot \pi^K ] ) +  \nonumber \\
&& \  2 z'_{0JK} [ \partial \dot \pi^J ] [ \partial \pi^K ] \bigg) + F_b \bigg( \frac{1}{2} \dot {\pi}^2 - \frac{1}{2} c_p^2 [\partial \pi]^2   - \chi^2 (\partial_\mu \dot\pi \cdot \partial^\mu \dot\pi  + [\partial\dot \pi \cdot \partial \dot \pi] ) \bigg),
\end{eqnarray}
\end{fleqn}
from the interacting  
\begin{fleqn}
\begin{eqnarray}\label{fdsa2}
\mathcal{L}_{\text{int}}  \supset &&   w_o \bigg[ \frac{c^2_s}{2} [\partial \pi] [ \partial \pi^2 ] + \frac{1}{2}( 1 + c_s^2)  [\partial \pi] \dot \pi^2  - \frac{1}{6} (3 c_s^2 + f_3)[\partial \pi]^3 - \dot \pi \cdot \partial \pi \cdot \dot \pi  - c_s^2 [\partial \pi] det \partial \pi \bigg] + \nonumber \\ 
&& w_o z_{IJK} \bigg[ \pi^I \cdot  \partial   \pi^J \cdot \partial \pi^{K} +  3 [ \partial \pi^I ] \pi^J \cdot \partial \pi^K  - \pi^I \cdot  \pi^J [ \partial \dot \pi^K ] -  ( \partial \pi^I ) \cdot ( \pi^J \cdot \partial \pi^K )  \bigg]  +    \nonumber \\
&& w_o \chi^2   \bigg[ [ \partial \pi] ( [\partial \dot \pi \cdot \partial \dot \pi ]   + (\partial_\mu \dot \pi) \cdot (\partial^\mu \dot \pi) )  \bigg] (  1 +  c_s^2 )  + z_{IJK} \bigg[ |\partial \pi^T \cdot \dot \pi^I |^2 \delta^J_K  + [ \partial \pi ]^2 \dot \pi^I \cdot \dot \pi^J \delta^K_0   \nonumber \\ 
&&  +  [ \partial \pi ] \dot \pi^I \cdot \partial \dot \pi^J \cdot \dot \pi^K + \ldots  \bigg] + z_{III} \chi^2   \bigg[ [\partial \dot \pi] [\partial \dot \pi \cdot \partial \dot \pi ] + \partial^2 [\partial^2 \pi ] \ddot \pi \cdot  \pi + 2 [\partial^2 \pi ]^2 [\partial \dot \pi ] + \ldots \bigg], 
\end{eqnarray}
\end{fleqn}
and self-interacting ones
\begin{fleqn}
\begin{eqnarray}\label{fdsa3}
\mathcal{L}_{\text{self-int}}  \supset &&  w_o^2 \bigg[ \frac{1}{2} [\partial \pi] \dot \pi^2 + \frac{c_s^2}{2} [\partial \pi]^3 + \frac{1}{2} [\partial \pi]^2 [ \partial \pi^2 ] + \frac{( 1 + c_s^2)}{2}  [\partial \pi]^2 \dot \pi^2  - \frac{1}{6} (3 c_s^2 + f_3)[\partial \pi]^4 -  [\partial \pi] \times   \nonumber \\ 
&& \dot \pi \cdot \partial \pi \cdot \dot \pi  -  c_s^2 [\partial \pi]^2 det \partial \pi \bigg] + z^2 \bigg[ ( \partial [\partial \pi ] ) \cdot (\partial [\partial \pi ]) [\partial \pi ] + 2 \dot \pi \cdot \partial \dot \pi [\partial \dot \pi ] + 2 [\partial \pi ] [ \partial \dot \pi ]^2 + \ldots \bigg]   \nonumber \\
&& + 2 \chi_b  \chi   \bigg[  [ \partial \pi]  [\partial \dot \pi \cdot \partial \dot \pi ] + [ \partial \pi ]  (\partial_\mu \dot \pi) \cdot (\partial^\mu \dot \pi)  \bigg] +  \chi \partial_{\omega^2} \chi \bigg[   (\partial_\mu \dot \pi^2 ) \cdot ( \partial_\mu \dot \pi^2 ) + 2 (\partial_\mu \dot \pi ) (\partial_\mu \dot \pi ) \times \nonumber \\
&& [\partial \dot \pi \cdot \partial \dot \pi ] + [ \partial \dot \pi \cdot \partial \dot \pi ]^2 + \ldots \bigg].
\end{eqnarray}
\end{fleqn}
To derive the Feynman diagrams, we need to present $W (J)$ \eqref{gen:func:hydro} in a manageable form. The main Feynman diagrams are in the \cref{tab:stimuli}. For the general scattering process, all amplitude is calculated in a long-wavelength approximation. The Feynman rules: all "in-" and "out-" states are \textit{on-shell}, it means, they satisfy the Euler hydrodynamics equation, internal line $1/w_o$, external line $1/\sqrt{w_o}$\footnote{A pedagogical review could be found in \cite{hydro0}.}. Consider the amplitude decay mechanism of the diagram below 
 
\begin{figure}[H]
\centering 
% of the shear stress tensor,
\includegraphics[width=3cm]{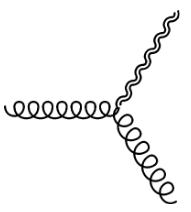}
\end{figure}

% \begin{fmffile}{simple_box}
% \begin{equation}
% \begin{fmfgraph*}(100,100)
% \fmfleft{i1}
%     \fmfright{o1,o2}
%     \fmf{gluon}{i1,w1}
%     \fmf{gluon}{w1,o1}
%     \fmf{dbl_wiggly}{w1,o2} 
% \end{fmfgraph*} 
% % \qquad \qquad
% % \begin{fmfgraph*}(90,90)
% % \fmfleft{i1,i2}
% % \fmfright{o1,o2}
% % \fmf{gluon}{i1,v1}
% % \fmf{gluon}{o1,v1}
% % \fmf{boson}{i2,v1}
% % \fmf{boson}{o2,v1}
% % \end{fmfgraph*}
% % \qquad \qquad
% \end{equation}
% \end{fmffile}
where a transverse polarization decay into a transverse polarization and excitation. The amplitude of this one tree-level diagram is 
\begin{equation}\begin{aligned}
i \mathcal{M_{T \rightarrow T T}} =& \frac{ (\hat{\epsilon} \cdot \hat{k}) \textbf{k}^2 }{\sqrt{w_o}} \bigg\{ z_{I00} [ 2    \sin{(3\theta/2)})] + z_{III,00} [     \sin{\theta}(c_s^2 - 2\cos{\theta}) ] - 4 c_s  \sin{(\theta/2)} + \\
&  z_{0I0} \chi^2 [ \omega \textbf{k}   \sin{\theta} + \textbf{k}^2 \sin{(2 \theta)} ]  \bigg\},
\end{aligned}\end{equation}
where $\theta$ is the angle between the ingoing and outgoing transverse mode. The kinematical restriction of conservation of energy and angular momentum, as well as the breaking symmetry, impose constraints on the amplitude. Before going further, we shall note that the energy required for the vortex to act as the emission of transverse excitations has to be greater than already found in \cite{DM}. If these specific conditions are satisfied, we can draw a class of relevant Feynman diagrams which manifest the coupling of $\eta$ with $\chi^2$. Now we restrict ourselves to three-level scattering problems $ \mathcal{TT} \rightarrow \mathcal{TT}$ where the polarized dissipative problems take new phenomena.

\begin{figure}[H]
\centering 
% of the shear stress tensor,
\includegraphics[width=3cm]{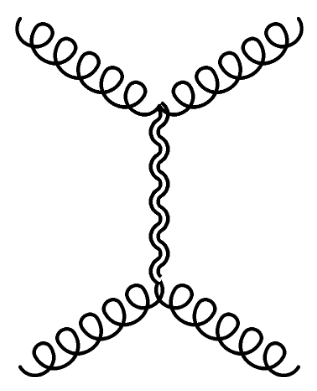}
 
\end{figure}

% \begin{fmffile}{boxxx}
% \begin{equation}
% \begin{fmfgraph*}(100,100)
% \fmfleft{i1,i2}
% \fmfright{o1,o2}
% \fmf{curly}{i1,v1,o1}
% \fmf{curly}{i2,v2,o2}
% \fmf{dbl_wiggly}{v1,v2}
% \end{fmfgraph*}
% \end{equation}
% \end{fmffile}
where the scattering amplitude for the diagrams above in the form 
\begin{eqnarray}
i \mathcal{M_{TT \rightarrow TT}} &=&  \frac{1}{w_o}  \bigg\{  z_{III} \chi^2 c_p^2 [ (\omega^2 \textbf{k}^4 - (\omega \textbf{k})^3 (3 - \cos{\theta})) + \omega^4 \textbf{k}^2 (\hat{k}_1 \cdot \hat{k}_2)] + z_{III} [  \omega^3 \textbf{k} + (\omega \textbf{k})^2 \cos{(\theta /2 )} ]  \nonumber \\ 
&& + \omega^4 ( 2 \textbf{k}^2 \partial_b \chi + \omega^3 \partial_{\omega^2} \chi )[(\hat{k}_1 \cdot \hat{k}_2) + \cos{\theta}]^2 + \frac{1}{2} \textbf{k}^4 c_s^2 (\cos{\theta} - \cos{( 2 \theta )}) \bigg\} i D^{(0)}_{ij} (\omega,\textbf{k}),
\end{eqnarray}
the terms proportional to $z_{III}$ contribute to the sound generation effect and  $\chi^2$ provides relevant information to the transverse generation by vortices source. These phenomena are qualitatively different from those studied in \cite{DM} because both the sound mode and the dissipative mode are included. Unavoidably, as we have seen above, this means that shear modes, not just propagating sound waves but the symmetric shear tensor fields sourcing heat via shear viscosity, couple to polarization via $\chi^2$ and shear viscosity. In the next sections, we shall explore the more direct physical consequences of this new process.

% \textcolor{red}{spin couplings cannot be neglected when the dissipative process are included. . we assume the two "sources" correlates. In other words, we can express as the product of two functions.  }

\subsection{Polarization mass correction}\label{pmc}
 
The case of virtual correction for ideal polarized fluid has been discussed in \cite{DM}. Our starting point is the free lagrangian \eqref{fdsa}, obtained by a linear perturbation the longitudinal polarization and excitation. 
\begin{equation} 
i H^{(0)}_{ij}  (\omega, \textbf{k} ) = \frac{1}{w_o} \frac{i L_{ij}}{\omega^2 - c_p^2 \textbf{k}^2 - \chi_0^{-2}}, \qquad \qquad i D^{(0)}_{ij}   (\omega, \textbf{k} ) = \frac{1}{w_o} \frac{i L_{ij}}{\omega^2-c_s^2 \textbf{k}^2 + i \gamma_\eta (z_0 )\omega \textbf{k}^2}, 
\end{equation}
where we denote the bare quantities with the subscript $0$. The beauty of this propagator is the presence of shear viscosity and spin variable. As we are primarily interested in the complete correction of the 2-point Green's function, we include the sum of all relevant one-particle-irreducible contribution
\begin{equation}\begin{aligned}
i H_{ij} (\omega, \textbf{k}) &= i H^{(0)}_{ij} (\omega, \textbf{k}) + i H^{(0)}_{im} (\omega, \textbf{k}) i \Tilde{\Sigma}_{mn} (\omega, \textbf{k})\, i H^{(0)}_{nj} (\omega, \textbf{k}) + \ldots, \\ 
i D_{ij} (\omega, \textbf{k}) &= i D^{(0)}_{ij} (\omega, \textbf{k}) + i D^{(0)}_{im} (\omega, \textbf{k}) i \Tilde{\Sigma}_{mn} (\omega, \textbf{k})\, i D^{(0)}_{nj} (\omega, \textbf{k}) + \ldots,
\end{aligned}\end{equation}
where the self-energy function $\Tilde{\Sigma}_{mn}$ encodes all reliable aspects of dissipative and polarized effects in effective field theory language. We begin with the trivial sum for the inverse propagator
\begin{equation} 
H_{ij}^{-1} (\omega,\textbf{k}) = \omega^2 - c_p^2 \textbf{k}^2 - \chi^{-2}_0 +  \Tilde{\Sigma}_{\omega,\textbf{k}}, \qquad \qquad  
D_{ij}^{-1} (\omega,\textbf{k}) = \omega^2 - c_s^2 \textbf{k}^2 + i \gamma_\eta (z_0 )\omega \textbf{k}^2 +  \Tilde{\Sigma}_{\omega,\textbf{k}},
\end{equation}
where $\Tilde{\Sigma}_{\omega,\textbf{k}}$ is the sum of all one-particle-irreducible Feynman diagrams. Since we have two cases to evaluate, we can decompose this self-energy contribution as
\begin{equation}\label{damped}
\Tilde{\Sigma}_{\omega,\textbf{k}} = i(\Sigma^P_{\omega,\textbf{k}} + \Sigma^S_{\omega,\textbf{k}} ) \textbf{1},
\end{equation}
being $\textbf{1}$ the unitary matrix. We label the polarized and shear contributions by the superscripts $S$ and $P$, respectively.  In the lowest order, the relevant contributions for one-loop diagrams are 
\begin{figure}[H]
\centering
\includegraphics[width=15cm]{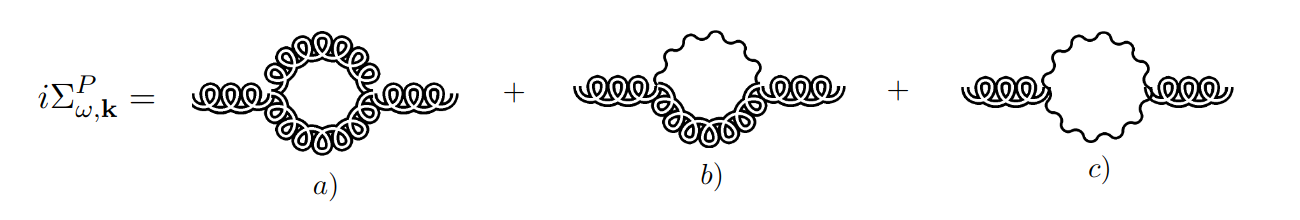}
\caption{The diagrams corresponds to the self-energy correction of the longitudinal polarization propagator.}
\label{loop:pol}
\end{figure}
Each loop in \cref{loop:pol} displays one general property of dissipative fluids with spin. Note that the $a)$ loop was already investigated in \cite{DM} corresponds to the dissipative vortex mass. The $b)$ loop tells us the interaction between shear and polarization is responsible for driving the vortex toward the principal axes and losing angular momentum from the vortex-coupling to the fluid. Finally, the loop $c)$ is mediated by the compressional modes. Since $\chi^2 Z_{III}$ in \eqref{fdsa2} includes the power counting $\partial^4$ higher than $w_o z_{III}$, It is not surprising that the non-renormalizable coupling $\chi^2$ turns the fluid dynamics into a higher derivative theory. In fact, such vortex configuration arises only if the momentum of internal line $k^2 > \chi^{-2}$. Below this gap, the vortices are "massless " low-energy degrees of freedom whose sound wave produced scatters elastically with the vortex-spin coupling. Above this gap, we have access to deep inelastic and absorption processes, and this short-range interaction in high energy leads to ultraviolet divergence since the vortex can assume an arbitrarily small size, which crosses the $\Lambda$.

Our effective field method requires a careful calculation to separate the infrared and ultraviolet parts, and as a consequence, the ultraviolet divergence breaks the $SO(3)$\footnote{The polarization Feynman propagator in \eqref{free:pl} are absent of ultraviolet divergences if we consider a relaxation time \cite{dm3}} symmetry. We can write the self-energy diagrams as   
\begin{equation}\begin{aligned} 
i\Sigma^P_{\omega,\textbf{k}} = - \chi^2 \int \frac{d^4 q}{(2 \pi)^4} H_{lm}^{(0)} (q) H_{mp}^{(0)} (k-q) &  - \chi z_{IJK}  \int \frac{d^4 q}{(2 \pi)^4} H_{lm}^{(0)} (q) D_{mp}^{(0)} (k-q) \\ &  \qquad \qquad \ \ 
 - z^2 \int \frac{d^4 q}{(2 \pi)^4}  D_{lm}^{(0)} (q) D_{mp}^{(0)} (k-q).
\end{aligned}\end{equation}
Let us now estimate the one-loop diagrams for the longitudinal excitation
\begin{figure}[H]
\centering 
% of the shear stress tensor,
\includegraphics[width=15cm]{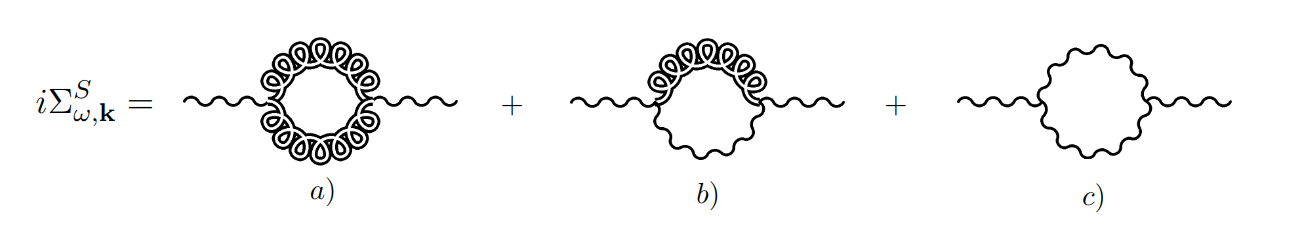}
\caption{The one-loop corrections to the longitudinal excitation.}
\label{loop:ns}
\end{figure}
The lowest diagrams for the longitudinal excitation in \cref{loop:ns} reflect the long-distance interaction of the fluid vorticose source. The $a)$ and $b)$ loops correct the speed of sound and $\gamma_\eta$ due to the interaction between fluid and spin. Moreover, they mediate the interaction between hydro-variable and vortex-coupling. In contrast with the general case where we observe the anisotropy due to the angle between spin and sound vector, we here restrict ourselves to sound waves propagating parallel to spin in \cref{loop:ns}, so the $c_s^2$ orientation is not anisotropy. We can compute these implications in the dynamics of longitudinal compressional modes by insertion of \eqref{damped} into $i D^{(0)}_{ij}$ [refer to Eq. \eqref{refere}].

The typical advantage of our framework is to absorb the ultraviolet divergence in the parameters of effective field theory. By linking the bare parameters with the measured ones, we renormalize the coupling and fields. The renormalization constants are expanding around tree-level solution 
\begin{equation}
Z_i = 1 + \sum_{j=1}^{\infty} \frac{1}{\epsilon^j}  Z_{ij} (z,\chi ) , 
\end{equation}
where the analytical function $Z_{ij}$ is independent of $\epsilon$ and exclusively dependent on hydrodynamic couplings. 
\begin{equation}
Z_{ij} = \begin{cases} 1 + \delta Z_{ii} , &  i =  j \\ \delta Z_{ij}, &  i \neq j \end{cases}    
\end{equation}
We split the bare fields and insert the renormalized constants to render ultraviolet finite states to Green's function \eqref{stctpform}. In the high-energy limit, we adopt the relevant independent parameters of hydrodynamics as effective field theory which extract the interaction of shear and spin. To observe physical quantities, the bare couplings absorb infinities   
\begin{align}\label{matrix_zx}
\begin{split}\begin{pmatrix}
\chi_0 \\ z^0_{IJK}
\end{pmatrix}
&=
\begin{pmatrix}
1 + \frac{1}{2}\delta Z_{\chi \chi} & \delta Z_{\chi z} \\
\delta Z_{ z \chi} & 1 + \frac{1}{2}\delta Z_{zz} 
\end{pmatrix}
\begin{pmatrix}
\chi \\ z_{IJK} 
\end{pmatrix}, \\
&\phantom{aaaaaa}
\end{split}\end{align}
where the renormalized parameters $\{ z_{IJK}, \chi \}$ are finite. The matrix of parameters is no longer diagonal. So we cannot express the renormalized parameters as eigenvalues of the bare ones. We renormalize the field $\pi$ independently  $\pi_0 = Z^{1/2}_{\pi \pi}=(1 + \frac{1}{2}\delta Z_{\pi \pi}) \pi$. The on-shell renormalization condition fixes the "mass", found in \cite{DM}, or minimal dissipative correction to the longitudinal polarization propagator $\Sigma^P_{\omega,\textbf{k}}  (k^2)|_{k^2 = \chi^{-2}} = 0$ and another renormalization condition is $\frac{d}{d k^2} \Sigma^P_{\omega,\textbf{k}} (k^2) \bigg|_{k^2 = \chi^{-2}} = 0$. Up to first order, the sum of all $1PI$ diagrams is  $\Sigma^P_{\omega,\textbf{k}} =  Z_{\pi \pi} +  \frac{k^2}{2} \delta Z_{\pi \pi} - ( \frac{1}{2} \delta Z_{\chi \chi} + \frac{1}{2} \delta Z_{\pi} + \delta Z_{z \chi} ) \chi^{-2}_R$, with the regularized mass $\chi^{-2}_R$. As expected from \eqref{prove:1} the assumption of microscopic reversibility, the shear and polarization dissipative currents take place simultaneously on a fluid cell. Switching these probe forces produces the same effect (commutation). Then the counterterms of non-diagonal elements obey the relation 
\begin{equation} 
\delta Z_{\chi z} = \delta Z_{z \chi}.    
\end{equation}
The detailed balance condition shows that transport coefficients are not dependent on the path history, but rather on their immediate predecessor states. Next, we write the exact propagator of longitudinal polarization and excitation as
\begin{equation}\label{refere} 
i H_{ij}  (\omega, \textbf{k} ) =\frac{1}{w_o} \frac{i L_{ij}}{\omega^2 - c_p^2 \textbf{k}^2 - \chi^{-2}}, \qquad \qquad i D_{ij}  (\omega, \textbf{k} )  = \frac{1}{w_o} \frac{i L_{ij} }{\omega^2-c_s^2 \textbf{k}^2 + i \gamma_\eta (z )\omega \textbf{k}^2 }. 
\end{equation}
where the renormalized coupling constant are $\chi^2 (\omega) = \chi_0^2 + Im \Sigma^S_{\omega, \textbf{k}} + Re \Sigma^P_{\omega, \textbf{k}} $ and       $\gamma_\eta (\omega) = \gamma_\eta (z_0) + Im \Sigma^P_{\omega, \textbf{k}} + Re \Sigma^S_{\omega, \textbf{k}}$. According to the Lagrangian \crefrange{fdsa}{fdsa3}, we evaluate the relativistic correction to shear and vortical susceptibility up to the first order. We suppress the cutoff by the renormalized couplings. The physical measurable quantities are 
\begin{eqnarray}\label{loop1}
\gamma_\eta (\omega) &=& \gamma_\eta    + \frac{23 T }{30 w_o \pi } \bigg[  \frac{|\omega|^{1/2}}{8   \gamma_\eta^{3/2}}  + \bigg(  \frac{\omega \gamma^{-1/2}_\eta}{2c_s^2} + \omega^{3/2} \gamma_\eta^{-5/2} \bigg)  \ln{\bigg(\frac{\Lambda^2}{\omega^2}\bigg)} + \bigg( \frac{1}{16} + \frac{1}{7 \gamma_\eta^2 \omega^2} \bigg)\Theta(\omega^2) \bigg] \nonumber \\ 
&& +  \frac{4 T^2 }{27 \pi^2} \bigg[ \frac{(\chi \gamma_\eta)^{1/2}}{w_o (\chi^{2} \omega^2 + 1 )^{2/3}} + \frac{2 \chi^{5/2}}{w_o \gamma_\eta^{1/2}} \bigg( \frac{|\omega|^{1/2} }{(2 + \chi^2 \omega^2)} \bigg)  + \bigg( \frac{1}{3} + \frac{\omega^2 \chi^2}{15}  \bigg) \sqrt{1 - \omega^2 \chi^2} \nonumber \\ 
&& \times \bigg( \frac{\gamma_\eta}{|\omega|}\bigg)^{1/2}  \Theta (\omega^2 - \chi^{-2}) \bigg] , \\ \label{loop2}
% &&&&&&&&&&&&&&&&&&&&&&&&&&&&&&&&&&&&&&&&&&&&&&&&&&&
\chi^2 (\omega) &=& \chi^2 +  \frac{T^2}{(4\pi)^2 w_o} \bigg( 1 + \frac{2}{3} \chi^2 + (6 + 2 \omega^2 \chi^2 + \frac{\chi^4}{4}) \ln{(\chi^2 \Lambda^2)} \bigg) + \frac{T^2 \chi^{3/2} |\omega^{1/2}| }{3\pi^2 w_o }\bigg( 4 - \frac{ \chi^{-2} - \mu^2}{\omega^2} \bigg)  \nonumber \\
&& \times  \sqrt{ \bigg( 1 - \frac{\chi^{-2} - \mu^2}{\omega^2} \bigg)  - 4 \chi^2 \mu^2 } + \frac{7 T^2}{25 \pi^2 w_o} \bigg(  1 + \frac{1}{6} \omega^2 \chi^2 \bigg) \ln{\bigg(2 - \frac{\omega^2 - \chi^{-2} }{\mu^2}\bigg)}  \nonumber \\ 
&& +  \frac{4 T^2 }{27 \pi^2} \bigg[ \frac{(\chi \gamma_\eta)^{1/2}}{w_o (\chi^{2} \omega^2 + 1 )^{2/3}} + \frac{2 \chi^{5/2}}{w_o \gamma_\eta^{1/2}} \bigg( \frac{|\omega|^{1/2} }{(2 + \chi^2 \omega^2)} \bigg) + \bigg( \frac{1}{3} + \frac{\omega^2 \chi^2}{15}  \bigg) \sqrt{1 - \omega^2 \chi^2} \nonumber \\ 
&& \times \bigg( \frac{\gamma_\eta}{|\omega|}\bigg)^{1/2} \Theta (\omega^2 - \chi^{-2}) \bigg].
\end{eqnarray}
We omit the subscript from $\{ \gamma_\eta^{ren} , \chi^2_{ren} \} $ to keep the notation light. This process reflects, in the lagrangian picture, the phenomenon identified in \cite{palermo,shearspin}. The symmetric shear and polarization states have the same symmetry properties. Hence, it is natural to expect them to mix. \cite{palermo,shearspin} characterize the mixing process as non-dissipative, and, indeed, the equations above make it clear that they are based on microscopic susceptibility and occur under conditions of detailed balance.

However, one should note that the symmetric shear is {\em not} an equilibrium quantity and generally relaxes to zero as global equilibrium is reached in a fluid, as it carries no conserved quantum numbers.  Polarization also relaxes, generally not to zero but to the {\em anti-} symmetric vorticity gradient carrying angular momentum \cite{dm3}.    The key here is the realization that under detailed balance the mixing happens instantaneously. This generally violates causality, as we showed before in \cite{dm3}. The next section explores how extending the action to second-order clarifies the relationship between the non-dissipative mixing of transient quantities, and also determines under what conditions this mixing really occurs.

\subsection{Second-order fluid action}
In the previous sections, we discuss the first-order corrections of hydrodynamics as an effective field theory. The introduction of this new language is still related with well-known ill-defined problems. This is because the dissipative processes in \crefrange{eqsshear}{eqspol} lead to acausality and instability effects in relativistic frame \cite{denicol2008ha}.

We indeed expect the introduction of correlations between microstates with no equilibrium counterpart since the deviation from the environment ($\Psi-$sector), at the local level, has no meaningful hydrodynamic approach. This assumption and local causality require \cite{DM} the inclusion of additional degrees of freedom breaking the symmetries associated with the equilibrium but having relaxational dynamics (although fluctuations mean these are not uniquely defined \cite{gavassino}). Microscopic interactions are responsible for the relaxation of degrees of freedom and the second law of thermodynamics means that in the absence of backreaction \cite{hydroryb} dynamics should be relaxational with respect to their source. This requirement, which turns out the Lagrangian of \crefrange{eqsshear}{eqspol} causal and stable, enlarges the parameter space via the introduction of new couplings. Hence, the new Lagrangian are 
\begin{align}\label{all:action2}
S &= S_{free} + S_{IS-shear} + S_{IS-bulk} + S_{IS-pol},  \\      
S_{IS-shear} &= \int d^4 x \bigg( \frac{\tau_\eta}{2}  \left({\pi^{\mu\nu}_-} u^\alpha_+ \partial_\alpha {\pi_{\mu\nu +}} - {\pi^{\mu\nu}_+} u^\alpha_- \partial_\alpha {\pi_{\mu\nu -}} \right) \ + \ \frac{{\pi^{\mu\nu}_{\pm}}^2}{2} \nonumber \\
 & \quad \quad \qquad \qquad \qquad \quad \qquad \quad + \ \ T^4_o   z_{IJK}(K^{\gamma} K_\gamma) b^2 B^{-1}_{ij} \partial^\mu \phi^{iI} \partial^\nu \phi^{jJ} \partial_\mu K^K_\nu   \bigg), \label{eqsshear2} \\
S_{IS-bulk} &= \int d^4 x \bigg( \frac{\tau_\xi}{2}  \left(\Pi_- u^\alpha_+ \partial_\alpha \Pi_{ +} -  \Pi_+ u^\alpha_- \partial_\alpha \Pi_{ -}\right) \ + \ \frac{\Pi_{\pm}^2}{2} \nonumber \\
& \qquad \qquad \qquad  \qquad \qquad \qquad \qquad \quad  \ \ \ + \ T^4_o    h_{IJK}(K^{\gamma} K_\gamma) K^{\mu I} K^{ \nu J} \partial_\mu K^K_\nu \bigg),   \\
S_{IS-pol} &= \int d^4 x \bigg( \frac{\tau_\chi}{2}  ( \, Y^{\mu\nu}_{-} \, u^{\alpha}_{+} \partial_\alpha Y_{\mu\nu +} - Y^{\mu\nu}_{+} u^{\alpha}_{-} \partial_\alpha Y_{\mu\nu -} ) + \frac{Y^{\mu\nu 2}_{\pm}}{2} + F (b ( 1-c y^2 )) \bigg). \label{eqspol2}
\end{align}
The on-shell equations of motion are
\begin{equation}\label{new:eom}
\tau_{\eta,\xi,\chi}  u_\alpha \partial^\alpha \left\{
\begin{array}{c}
\pi^{\mu \nu} \\
\Pi \\
Y^{\mu \nu}
\end{array} \right\} + \left\{
\begin{array}{c}
\pi^{\mu \nu} \\
\Pi \\
Y^{\mu \nu}
\end{array} \right\} =
\left\{
\begin{array}{c}
\Delta^{\mu \nu \alpha \beta} \partial_\alpha u_\beta\\
\Delta^{\alpha \beta} \partial_\alpha u_\beta\\ 
\chi \omega^{\mu \nu}
\end{array} \right\},
\end{equation}
where $ \Delta^{\mu \nu \alpha \beta} = \frac{1}{2}( \Delta^{\mu\alpha} \Delta^{\nu\beta} + \Delta^{\mu\beta} \Delta^{\nu\alpha} - \frac{2}{3}
\Delta^{\mu\nu}\Delta^{\alpha\beta}) $. The $\pi^{\mu\nu}$, $\Pi$, and $Y^{\mu\nu}$ are new dynamical variables, which relax to their first-order gradient expansion: shear $ \pi^{\mu\nu}_{|\textit{shear}}$, bulk $\Pi_{|\textit{bulk}}$, and polarization $ Y^{\mu\nu}_{|\textit{pol}}$ by following their respective characteristic time scale. In absence of homogeneous part, the fluid dynamic force decays exponentially to zero on the time scale $\tau_{\eta,\xi,\chi}$ dictated by the subjacent microscopic interactions. As per the Israel-Stewart prescription, the acausal modes into \crefrange{true1}{true3} must be crucially cut-off after replacing the transport coefficients    
\begin{equation} 
\left( \begin{array}{c}
\eta\\
\xi\\
\chi
\end{array}
\right) \rightarrow
\left( \begin{array}{c}
\eta\\
\xi\\
\chi
\end{array}
\right) \frac{1}{1+ i \omega \tau_{\eta,\xi,\chi}}.
\end{equation}
The new Green-Kubo formulae are

\begin{eqnarray} 
&& \frac{1}{2} \lim_{\omega \to 0} \lim_{k \to 0} \partial^2_\omega ImG^{R}_{J^{z},\omega^{xy}} (\omega,\textbf{k}) = \chi^2 \tau_\chi, \\
&& \frac{1}{2} \lim_{\omega \to 0} \lim_{k \to 0} \partial^2_k ImG^{R}_{J^{z},T^{xy}} (\omega,\textbf{k}) =
\tau_\eta ( \Bar{z}_{333} + \Bar{z}_{303} + \Bar{z}_{033} + \Bar{z}_{003}) + \tau_\xi ((\bar{h}_{003,03} + \bar{h}_{303,03} + \bar{h}_{333,03} + \nonumber \\ 
&& \bar{h}_{033,03} -2 \bar{h}_{333,03} -2\bar{h}_{303,03} +2 \bar{h}_{330,03} +2 \bar{h}_{300,03}) - 4 b_o^5 (\bar{h}_{333,00} + \bar{h_{303,00}}) + 4 b_o^5 ( \bar{h}_{003,33} + \bar{h}_{303,33}  \nonumber  \\ 
&& +  \bar{h}_{333,33} + \bar{h}_{033,33} + \bar{h}_{330,33} + \bar{h}_{300,33}) - b_o^{3}(\bar{h}_{003} + \bar{h}_{303} +   2\bar{h}_{333} + 2\bar{h}_{033} + \bar{h}_{333} + \bar{h}_{300} )). 
\end{eqnarray} 
Our main goal is to restore the causal behavior of fluctuations encoded in the kernel of \eqref{old:GK}. In obtaining these relations, we can determine $ \{ \tau_\chi ,\tau_\eta,\tau_\xi, \chi, \eta, \xi \}$ from thermodynamical identities and conserved quantity. This is a good illustration of the fact that $\tau_\chi$, $\tau_\eta$, and $ \tau_\chi$ (from \crefrange{eqsshear2}{eqspol2}), even in a ``bottom-up'' effective theory are not arbitrary but reflect how different fluctuations conspire. However, as shown in \cite{hydro1}, there is no well-defined limit when $\eta,\chi,\tau_\eta, \tau_\chi$ go to zero without IR instabilities due to vorticity.

The new set of variables in \eqref{new:eom} arises non-hydrodynamics modes, obeying the causality and stability conditions. As an approach to studying the fluid behavior in high $( \omega , \textbf{k})$, these collective modes help to extend, outside of the hydrodynamic regime, our analysis by keeping track of microscopic processes. For $t>0$, we evaluate the tensor-tensor correlation function \eqref{billy} with the inclusion of the relaxation time
\begin{equation}\begin{aligned}\label{last:one}  
\langle \{ T^{ij}  (t,\textbf{x}) , T^{kl} (0) \} \rangle  \sim &  \mathcal{O} (\Lambda) + H^{ij}_{kl} \frac{T^2}{6 \pi} \bigg[  \frac{\chi^{-5/2}}{ w_o ((1 + \tau_\chi^2 / \chi^2)t)^{3/2} } + \bigg( 3 + \bigg( \frac{1}{6} \bigg)^{1/2} \bigg) \frac{\chi^{-3/2}}{2((1 + \tau_\chi^2 / \chi^2) t)^{1/2}  } \bigg]  +  \\
&  \bigg( 7 + \bigg( \frac{3}{2} \bigg)^{3/2} \bigg)  \frac{S^{ij}_{kl} }{60 \pi} \frac{T^2}{ ( \gamma_\eta (z) t)^{3/2}} + \ \frac{H^{ij}_{kl} T^2}{ ( ( \gamma_\eta (z) + \frac{4}{3} \gamma_\xi (z)) t)^{3/2}} \ +  \frac{T^2}{2 \pi w_o} \bigg[ \frac{e^{ - \gamma_\eta \textbf{k}^2 t}}{32 }  \\ 
&  +  \frac{1}{3} k^i k^j e^{- \frac{1}{2}\textbf{k}^2 \gamma (z,h) t} \cos{(|\textbf{k}| c_s t)}  \bigg] \frac{ \bar{ \langle  p^2 \rangle } }{\mathcal{V}} + \frac{T^2}{2 \pi} e^{- \chi^2 \textbf{k}^4 t / f_b b } \frac{ \Bar{\langle y^2 \rangle} }{\mathcal{V}} + \ldots.
\end{aligned}\end{equation}
This generalized matrix corresponds to transverse  conserved quantity from the stress dynamics. One essential feature is that the polarization dynamics satisfy the bounded causality relation $ \frac{d P}{ds} \leq  \frac{\chi^2}{\tau^2_\chi}$ \cite{dm3}. The manifestation of oscillatory solution concerns to the longitudinal momentum solution, irrelevant for long-time tails. The pole solution for small wavenumber of \eqref{eq:state-alice-antirob} is 
\begin{equation}\label{last:two}
\omega (\textbf{k}) \simeq - \frac{i}{\tau_\eta} - \frac{i}{\tau_\chi} + i\gamma_\eta ( z ) \textbf{k}^2  + i   \frac{\chi^2}{f_b b}  \textbf{k}^4 + \ldots.
\end{equation}
This dispersion relation, partially in accord with \cite{kovtail}, rules the system evolution at early time when the scales gradient expansion scale is  weaker than non-hydrodynamics modes. Instead of hydrodynamics modes which survive for $k \rightarrow 0$, the decaying of collective excitations evolves according to $Re[\omega (\textbf{k} \rightarrow 0)]=$ finite damping terms. Note that the hydrodynamic modes exhibit an infinite lifetime for long-wavelength limit, while the lifetime of collective modes, characterized by the inverse of damping behavior, increases with $\textbf{k}$.

We identify the high wavenumber dependence of $\chi$ as an aspect denoting how fast spin increases in non-hydrodynamic regime with an exponential decay time $ \sim e^{-\omega (\textbf{k}) t}$ with $Re [\omega (\textbf{k})] \propto \textbf{k}^4$, while the shear with $Re [\omega (\textbf{k})] \propto \textbf{k}^2$. Indeed, a more careful analysis of the relaxation time opens the door for different qualitative results since this calculation is affected by nonlocal in time contribution, see discussion in \cref{mef}. Such statement, largely studied in the literature \cite{jepeux}, requires special attention because it is related to the critical wavenumber that each transversal mode propagate.

We explicitly recall that, in rotational dissipative fluid, the transverse momentum \eqref{for:2} is nontrivial and can be divided into two parts: spin and hydro-particle. The eigenvalues of stress tensor operator \eqref{last:one} express, on the microscopic scale, the propagation of shear viscosity and transverse polarization from kinetic modes. Using hydrodynamic limit $\textbf{k} \rightarrow 0$, one can deduce restrictions based on the kinetic characteristic of Maxwell relaxation time \cite{mrt} to illustrate the emergence of these collective modes. We can go further, at this point, and determine a wavenumber limit in which fluid shows dynamics of transverse excitations. For such an interpretation, the distance $L \sim 2 \pi / \textbf{k}$ is the upper bound limit to observe these excitations. Following the procedure in \cite{mrt2}, we find   
\begin{equation}\label{oublier}
\textbf{k}_\eta \sim \bigg[ \frac{ \epsilon G_\eta }{ \eta^2} \bigg]^{1/2}_{\textbf{k} \rightarrow 0}, \qquad \textbf{k}_\chi \sim \bigg[ \frac{ f_b b }{8 G_\chi \chi^4} \bigg]^{1/4}_{\textbf{k} \rightarrow 0}.
\end{equation}
The shear-stress $\textbf{k}_\eta $ and transverse polarization $\textbf{k}_\chi$ modes show wave-solutions propagation for $\textbf{k} > \textbf{k}_\eta $ and $\textbf{k}_\chi$\footnote{ $\tau_\eta = \eta / G_\eta  $ and $\tau_\chi = \chi^2  G_\chi $ with higher frequency transversal shear and polarization modes $G_\eta$ and $G_\chi$, respectively.}, whereas these transverse dynamics present diffusive behavior for length $ > L_\eta $ and $ L_\chi$. This last interpretation is, in fact, reinforced by conservation laws in hydrodynamics picture. It means the balance equation of transverse momentum $p_T$ and longitudinal one $p_L$ are decoupled. Hence, the $p_T$ dynamics is purely diffusive, and its macroscopic current must be limited to the shear relaxation process $\tau_\eta$. 

Upon comparing $L_\eta$ and $ L_\chi$, the rotational fluid arises a striking difference because the excitation of collective modes depends on how $\eta$ and $\chi$ compete. These qualitative and quantitative characteristics discuss the corrections, induced by thermal fluctuations, for spin and shear viscosity when the extrapolation of the domain spectrum approaches interatomic distance (high frequency). In the case of $L_\eta > L_\chi$, the shear waves have a dominant role in the collective modes before the manifestation of wave-like transverse polarization. The former corrects positively the shear viscosity by transferring $(x)$momentum in the $(-y)$direction \cref{viscfig}. The extension of this result permits to increase more spin displacement towards the principal axes of rotational fluid. As a result, it should be included in the future simulation data to track the momentum distribution \cite{cans}. On the other hand, $L_\eta < L_\chi$ first manifests the collective modes of transverse polarization, which corrects the $G^R_{T^{0i} T^{0j}} ( \omega , \textbf{k} )$ by $\langle J_P^T (t, \textbf{k} ) J_P^T ( t=0, \textbf{k}) \rangle$, with $J_P^T$ the polarization microscopic current of transverse momentum. Our main results are encoded in \eqref{last:one}, higher-order terms of \eqref{tensordef} will lead to subdominant effects in the long-time tails with time power-law of $t^{5/2}$ and $t^{7/2}$.

It turns out that after increasing $\textbf{k}$ to the region $ < L_\eta  $ and $ < L_\chi $, both of transverse modes are relevant for correcting the transport coefficients which turn into the frequency dependence $\{ \gamma_\eta (\omega),  \chi (\omega) \}$ and reduce to the "bare" quantities $\{ \gamma_{\eta} , \chi \}$ in $lim_{\omega \to 0}$. In this region, the new correlation function couples the microscopic transverse current of shear $J^T$ and $J_P^T$ as $\langle J^T (t, \textbf{k} ) J_P^T ( t=0, \textbf{k}) \rangle$, which contributes to $G_{T^{0i}J^i}^R (\omega , \textbf{k} = 0)$ in \eqref{old:pol}. However, this correlation is not equal to $\langle J_P^T (t, \textbf{k}) J^T (t=0,\textbf{k})\rangle$ because of broken Onsager symmetry. As the dissipative processes depend on the previous history, this propagator cannot be constructed in a linear way $G_{T^{0i}J^i}^R (n t) = [G_{T^{0i}J^i}^R (t)]^n$. Then the previous steps of microscopic distribution derive distinct time correlation functions. To understand what is happening in an aged system, the preparation of the propagator yields different macrodistribution from the same stationary state.

Many hydrodynamics treatments are build-up for low $(\omega,\textbf{k})$. Hence, they cannot provide a precision satisfactory for high $\omega$ by a straightforward extrapolation for higher derivatives. That is why the results \eqref{oublier} are important to understand the relevant aspects of fluid dynamics. In our approach, the magnitude of $\textbf{k}_\eta $ and $\textbf{k}_\chi$ cannot be determined by analytical methods, as it is done in \cite{mrt2}. We can only make progress in investigating further questions by appealing to experimental data. In addition, we are not able to distinguish which wave occurs first since the collective excitation of the shear and polarization, originated from different microscopic forces, depend on the spatial scale. We will postpone the analysis of the transport coefficient $\{ \eta (\omega), \xi (\omega), \chi (\omega) \}$ in future work.

\section{Memory effect}\label{mef}

The presence of spin in perfect fluid creates inherently an out-of-equilibrium scenario \cite{Hattori}. The complexity of such a system concerns the incorporation of micro- and macro-scales. Consequently, the theoretical difficulty increases as long as experimental possibilities. As we can always cover the same phenomena from "bottom" to "up" by different coarse-grained processes, the strategy to set the appropriate hierarchical levels is regarding how far hydrodynamics fluctuations are from linearized hydrodynamics. Nonetheless, what we call “ideal spin hydrodynamics” must be dissipative by construction \cite{dm2}. One of the reasons is the absence of symmetry conserving separately spin and angular momentum demands a new microscopic degree of freedom. We already saw the relevant aspects for a more accurate description of fluid dynamics are related to the high energy spectrum region. To achieve such task, we shall withdraw unnecessary variables to turn correlation functions ultraviolet finite since this measurable quantity comes from simulation data and experiments. 
% for the case of spacetime evolution of spin density,

In relativistic nuclear collisions, hydrodynamics models validate the interpretation of data in the QGP as a quasi-ideal fluid \cite{Heinz}. Strictly speaking, the hydro-signature reproduces wealth information about the QGP because we assume a local thermal equilibrium to the collective flow. However, most proposal experiments to measure the spin polarization and understand the hydrodynamical prediction run into conceptual issues. One conceptual issue involves technical problem of pseudo gauge invariance to the extension of thermodynamical corrections beyond the local equilibrium. We shall remind the theory behind the spin dependence on momentum \cite{lisa} shows disagreement when contrasted with experiments \cite{newone}. There are many methods to discuss this open problem in hydrodynamics, and each scenario provides deep insight into reducing such discrepancy (for review see \cite{Becattini2022zvf,kodi}). To improve hydrodynamics as a reasonable model to study the rotation of heavy ion collisions, we shall include, in the transport process, the interaction effects between shear and polarization. The dissipative forces as shear $ \sim l_{mfp}/ l_{hydro} $ and polarization $\sim l_{micro} / l_{mfp}$ induce new phenomena to non-equilibrium macroscopic dynamics. Even though its relation may appear rather complicated, the identification of the rightful coefficients of dissipative spin current might cure the discrepancy found at high and low energy \cite{qwer,shearspin}. The arguments raised by the Markov analysis, so far, are well suited for near-equilibrium approximation. However, we are left with the case of nonlinear transport of hydrodynamical variables in \eqref{last:one} and \eqref{old:non}.   

% \footnote{see discussion for massless fermion \cite{asdf}}

The aforementioned \cref{chap:3} and \cref{chap:4} show fluctuation-dissipation theorem became an indispensable tool to describe the coupling of shear and polarization. To do this, we assume reversibility temporal and time-symmetry as the majors rule at the microscopic scale. As we have seen, the cross-phenomena between $\eta$ and $\chi$ stands, within a certain approximation, because of the Markov characteristic. By comparing \eqref{old:pol} and \eqref{for:2}, whenever the transfer of angular momentum happens, the displacement of spin towards the rotational axes follows $\delta t$ time later. On the other hand, the inverse process must occur at the same average rate with an odd external vorticity field $\Omega$. Hence, since the fluid does not depend on the previous history, the correlation function for $p^i_T$ and $Y^{\mu \nu}$ in \eqref{old:GK} assume the form of 
\begin{equation}\label{prove:1}
\langle p^i_T (t), Y^{\mu\nu} (t+ \delta t,\Omega) \rangle_{eq.} = \langle Y^{\mu\nu} (t,- \Omega) , p^i_T (t+ \delta t) \rangle_{eq.}
\end{equation}
The Onsager symmetry shows that the approximation for the linear perturbation minimizes the memory effect. Indeed, this irreversible current demands the scale time of spin evolution is comparable with the hydro-variables in the near-equilibrium state. Thus, the microscopic interactions yield the same correlation function for both processes and demand a special connection between $\eta$ and $\chi$. In fact, the generator of dynamics leads to the same physical state that the generator of time-reversed symmetry.

By looking at the conservation of stress tensor up to the first order $\partial_\mu \langle T^{\mu\nu} \rangle = \langle \partial_\mu  T^{\mu\nu} \rangle = 0$, Noether's theorem is another reason to adopt the Markovian approximation. This relation arises naturally because the collective behavior can be generalized to a continuous medium, and the interaction between fluid variables and the background configuration can be neglected. In what follows, fluctuations are not observed in the $l_{hydro}$ scale, and stochastic variables are present in linear equations as white noise. For that reason, the environment restores its stationary solution since the perturbation of hydro-variables does not survive for a long-time. One of the insights is to compare the relaxation time of hydro-variables-$t$ and thermo-fluctuations-$t_m$, what one calls "system" and "bath", respectively. In case the former time scale is larger than the latter $t \gg t_m$, we can make a slow approximation, and so the average becomes $ \langle \pi (x) \rangle = 0 $ and correlation $ \langle \pi (x_1) \pi (x_2) \rangle = C_{12} \delta (x_1 - x_2)$, using the shorthand convention $x \equiv (t_1, \textbf{x}_1)  $ with $C$ being the matrix of coefficients determined by \eqref{old:GK} and conservation laws.

Under the assumption of the Onsager hypothesis, the investigation of effective field theory creates a scenario in which the one-loop corrections in \eqref{loop:pol} and \eqref{loop:ns} are simultaneous. We can then write down the counterterms $\delta Z_{\chi z} = \delta Z_{z \chi}$ produce an orthogonal matrix in \eqref{matrix_zx}. The corrections in \eqref{loop1} and \eqref{loop2} lose their applicability in effective field theory when the relaxation process takes place.

Having examined the fluid with spin under rotation, we will explore general features of the non-markovian properties in the transport coefficient. At a coarse level, the fluid under certain physical circumstances can lead to a process where the memory effects influence the macroscopic dynamics \cite{palermo2}. One of them is when, under certain circumstances, the relaxation time scale of hydrodynamical and non-hydrodynamical variables are comparable \cite{possibl}. The need for a non-markovian process may not seem obvious when theories beyond local equilibrium dismiss that the relaxation time of microscopic variables reflect the macroscopic dynamics. When the hydrodynamics fluctuations are within the second order dissipative equation, the fluid with spin faces memory effect. With the purpose to cure the discrepancy of experimental observable in QGP, we shall take into account the influence of past events of the system in the actual state.
%   Recalling the Onsager hypothesise lead to instantaneous response as approximation to steady state.  
% While the first-order hydrodynamics suits well the Markovian evolution, the Israel-Stewart to rotating fluid 
% Because of the minimum energy principle, the fluid assumes an anisotropic configuration with an ellipsoid shape.

\begin{figure}[H]
\centering
\includegraphics[width=16cm]{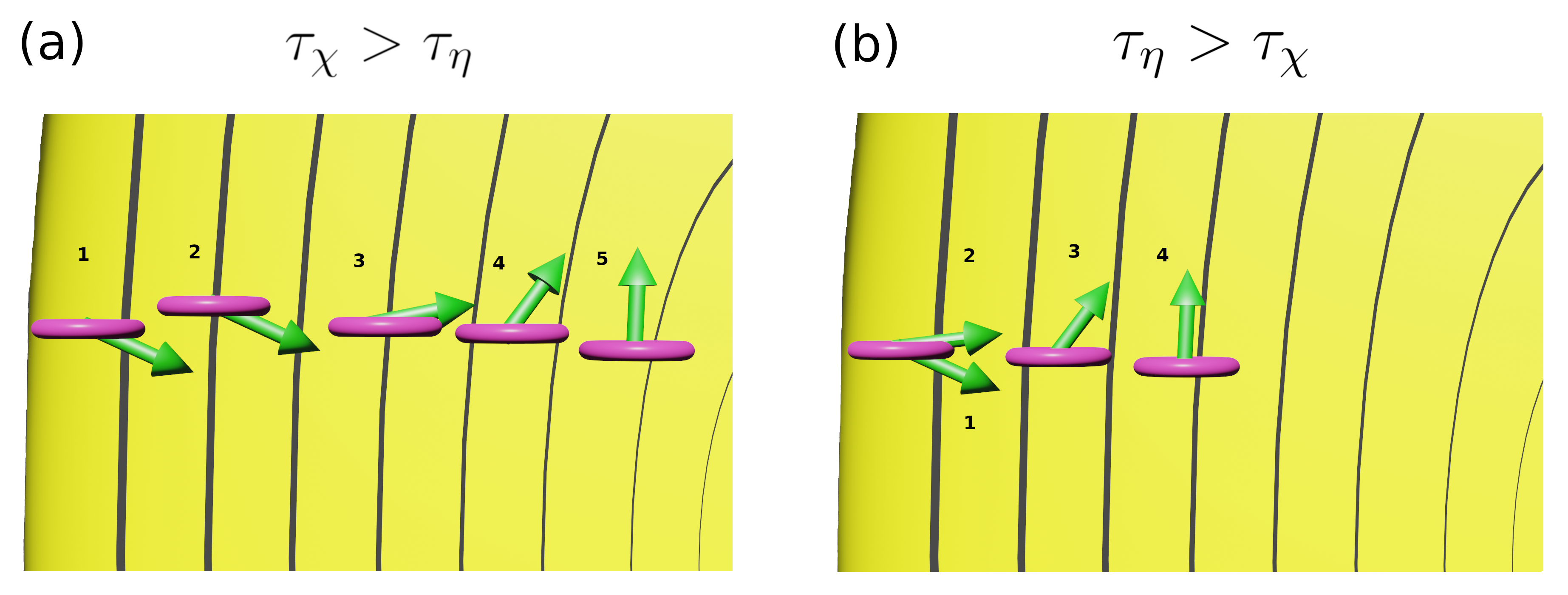}
\caption{This schematic picture shows how the shear inclusion redistributes the spin location in the fluid. The displacement towards the rotation axis depends on the relaxation time characteristics.}
\label{fig:galaxy2}
\end{figure}

Since the $\chi$ and $\eta$ have different symmetries \eqref{table:ss}, the rate of mechanical relaxation must provide two different scenarios in hydrodynamics. By keeping in mind \eqref{new:eom}, the relaxation time $\{ \tau_\eta, \tau_\chi \}$ are crucial to understand this cross effect. We also expect the specification of symmetric and antisymmetric inter-molecular interaction yields different dynamics for the spin particle. To discuss both scenarios, let us first consider the spin out of equilibrium (not totally aligned with the external vortical fluid). For the first case $\tau_\chi > \tau_\eta$, the transfer of angular momentum from spin to fluid happens at $\tau_\eta$ time, and so the spin moves in direction to inner layers: points 1-2 in \cref{fig:galaxy2}. The spin alignment towards the external vortex direction only follows $\tau_\chi - \tau_\eta$ later: points 2-5 in \cref{fig:galaxy2}. On other hand, for the case $\tau_\eta > \tau_\chi $, the flow produces a scenario where the polarization alignment occurs before the spin moves towards the principal rotation axis. In this inverse process, whenever the spin begins to align at $\tau_\chi $ time: points 1-2 in \cref{fig:galaxy2}, its shifts towards the rotational axe only follows $\tau_\eta - \tau_\chi$ later: points 2-4 \cref{fig:galaxy2}, which is damped out more by viscosity at $\tau_\eta$ time. In both situations, the fluid reaches a thermodynamical equilibrium when the polarization current is parallel to the external vorticity field. Other scenarios are possible: $\tau_\eta \ll \tau_\chi$, $\tau_\eta \gg \tau_\chi$, or $\tau_\eta \sim \tau_\chi$. The classification of all the cases leads to a clear meaning of how to deal with the interaction between spin and fluid particles. The manifestation of non-Markov properties become evident if we examine the spectrum properties of the physical variables. This calculation yields significant knowledge of the out-equilibrium process since each relaxation time depends upon the nature of interaction: range-action, symmetry, potential, intermolecular-force, and so on.    
Taking the evolution time of density matrix in the interaction picture by \eqref{take}, we have 
\begin{equation}\label{cacete}
\rho_I (t)= e^{i \mathcal{E}_0 t}\rho_{eq}(t)e^{-i\mathcal{E}_0 t},\quad \mathcal{E}_0 =\mathcal{E}_0 (\phi) + \mathcal{E}_0 (\Psi).
\end{equation}
Showing that we can rewrite in a compact form
\begin{equation}\label{fdgs}
\dot{\rho}_I (t) = i [ \rho_I (t), \mathcal{E}_I (t) ],
\end{equation}
where the interaction Hamiltonian, coming from the interacting Lagrangian in \eqref{mix::ctp} evolves as $\mathcal{E}_I (t) = e^{i \mathcal{E}_0 t} \mathcal{E}_I (0) e^{-i \mathcal{E}_0 t} $.
% It is important to stress the fields composing the Lagrangian in \eqref{all:action2} evolves as the free ones. 
% The primary discussion in our non-Markov effects lies for a hierarchy of relaxation times in the phenomenological description. As in section \cref{flueft}  
The factorization $\rho_I = \rho_\phi \bigotimes \rho_\Psi $ and $Tr\{ \mathcal{E}_I (t) \rho_I (0) \} = 0$ are fundamental assumptions to implement a Markovian approximation. However, the inclusion of spin variables turn the system out-of-equilibrium whose microscopic time scale is comparable with the system for second order relativistic hydrodynamics. The general solution of \eqref{cacete} is $ \rho_I (t) = \rho (0) + i \int^t_0 d t^\prime  [ \rho_I (t^\prime), \mathcal{E}_I (t) ] $. An interactive recursive solution with \eqref{fdgs} leads to 
\begin{equation} 
\dot{\rho}_I (t) =  i [ \rho_I (0), \mathcal{E}_I (t) ]  +  \int^t_0 d t^\prime  [ [ \mathcal{E}_I (t^\prime) , \rho_I (t^\prime) ] , \mathcal{E}_I (t) ].  
\end{equation} 
The second interactive term is a noncommutative dissipative process  
\begin{equation}
\int^{t}_{t_0} d \tau_\chi \mathcal{E}_I (\tau_\chi) \int^{\tau_\chi}_{t_0} d\tau_\eta \mathcal{E}_I (\tau_\eta) \int^{\tau_\eta}_{t_0}  dt' \mathcal{E}_I (t') \neq \int^{t}_{t_0} d\tau_\eta \mathcal{E}_I (\tau_\eta) \int^{\tau_\eta}_{t_0} d \tau_\chi \mathcal{E}_I (\tau_\chi) \int^{\tau_\chi}_{t_0}  dt' \mathcal{E}_I (t'),
\end{equation}
where the $LHS$ and $RHS$ correspond to the left and right picture of \eqref{fig:galaxy2}, respectively. The presence of non-commutative diffusion dictates $\delta Z_{\chi z} \neq \delta Z_{z \chi}$ in effective field theory language. In particular, it is shown that the memory effects influence loop corrections. It is seen that the $a)$ correction of \eqref{loop:pol} and \eqref{loop:ns} occurs first, and so the information of this early states is transferred to the other ones at $\tau_\eta - \tau_\chi$ time. On the other hand, the $c)$ correction of \eqref{loop:pol} and \eqref{loop:ns} arises at $\tau_\eta$ time, while the other ones at $\tau_\chi - \tau_\eta$ time.

The friction, including memory effect, is not instantaneous but depends on the previous steps. These characteristics appear when the time correlation between microstates is not neglected, and so $\partial_\mu \langle T^{\mu\nu} \rangle \neq \langle \partial_\mu T^{\mu\nu} \rangle$. We can only assure this equality if freezing the environment for the short time in which non-hydrodynamics modes are relevant. Consequently, the fluctuations turn the colored noise degrees of freedom into white ones.

The aforementioned discussion opens a window of theoretical effort to deal with the discrepancy found in the $QGP$. We will see in the future publication of non-markovian effects are appropriate adjectives.

% The evolution time of the system $T_s$ depends on which time scale (relaxations pole $\tau$) the environment restores observable average from non-equilibrium to Gaussian distribution probability $\tau \ll T_s$.

\section{Summary and Outlook}\label{sao}
In this work we have examined lagrangian hydrodynamics with both polarization and dissipative effects, from the perspective of the field theory. We have examined the interplay of viscosity, vortical susceptibility and sound wave backreaction on fluid dynamics. We hope that this is a step towards a full theory of hydrodynamics with spin.

Our main conclusion is that the shear forces and the polarization dynamics ``do not commute'', resulting in several different regimes determined by the respective relaxation time-scales.    While this paper is theoretical and specific to the lagrangian picture and the linear response approach, it is directly related to different topics, both theoretical and phenomenological which have been discussed in the literature.
The fact that viscous forces interact with polarization has been realized in the context of coarse-graining Zubarev hydrodynamics \cite{becshear} and trasnport theory \cite{shearspin}.  It has been advocated as a solution of the the longitudinal polarization phenomenological puzzle \cite{Adam,becshear2,yinshear2}.   However, since the symmetric shear is not an equilibrium nor a conserved quantity, a general effective theory of this dynamics was missing.   This work has clarified the regime where such terms are significant in the scale expansion.

If \cite{becshear2,yinshear2} will become accepted as an explanation of the longitudinal spin puzzle, it would imply that the spin relaxation time is in fact not small w.r.t. the relaxation time of shear quantities, since,as we show, this is the regime where the shear forces drive polarization. It would likely mean that spin and vorticity are not in equilibrium throghout hydrodynamic evolution, and the Cooper-Frye type freezeout assuming it which has so far been used for phenomenology \cite{zanna} needs correcting. 

The ``mass correction'' derived in \cref{pmc} is equally interesting phenomenologically. It would mean vorticity is not linearly proportional to angular momentum but aquires components dependent on characteristic vortex size $\omega$ weighted by microscopic parameters ($\chi^2$ and $\gamma (z,h)$).  This is again of potential phenomenological interest in the transverse polarization (produced at the scale comparable to the system size) vs longitudinal polarization (produced on finer scales determined by anisotropic flow), as well as the impact parameter dependence of global polarization.    One would need to input a frequency dependent polarization susceptibility correction into the freeze-out code (in practice, a correction to the Boltzmann polarization factor depending on the vortex size) to estimate such effects quantitatively.    A conclusive deviation of the linear dependence of polarization with respect to impact parameter (not seen as yet \cite{lisa}) could provide evidence for such ``anomalous vorticity propagation''.

On the theory side, while the approach preseted here is based on the fluctuation-dissipation theorem, the backreaction on hydrodynamic evolution of fluctuations has not yet been explored.  It is reasonable that, analogously for \cite{Kovtun2012rj} spin fluctuations will affect evolution more in a regime where $\tau_Y$ is sub-dominant, while hydrodynamic fluctuations become important when shear relaxation time is small.   A full understanding of fluctuations is necessary if spin is to drive a ``ferromagnetic type'' (or rather ``verrovortetic'') phase transition, as originally discussed in \cite{dm3}.   The presence of a phase transition will add another scale, to be studied using the Landau theory of phase transitions \cite{dm3,Landau} and either  Maxwell construction or nucleation, depending on fluctuation probabilities. Large scale vortical structure, dissipation, fluctuation and phase structure would then each has a dominant regime.

In conclusion, we have developed an effective theory of dissipative hydrodynamics with spin, based on the lagrangian picture and linear response theory.  We hope this is a step towards the still elusive goal to understand, both at a theoretical and phenomenological level, the effect that spin dynamics has on hydrodynamic evolution.

% \section*{Aknowledgments}

\appendix
\section{Derivation of Green-Kubo Relations}\label{gk}
In this appendix, we briefly demonstrate the derivation of the Green functions from the higher-order stress tensor. The main object to accomplish this task is the Lagrangian which describes our rotational dissipative fluid system in \eqref{all:action}. Thus, we can expand the stress tensor for higher orders defined as

\begin{align}\label{st}
T^{\mu\nu} =& \bigg\{ \frac{ \partial \mathcal{L}   }{\partial (\partial_\mu \phi^I )  } - \partial_\beta \frac{ \partial \mathcal{L}  }{  \partial (\partial_\mu \partial_\beta  \phi^I )  } +  \partial_\beta  \partial_\gamma \frac{ \partial \mathcal{L}  }{\partial (\partial_\mu \partial_\beta  \partial_\gamma \phi^I )  } - \dots  \bigg]   \partial^\nu \phi^I +  \bigg\{  \frac{ \partial  \mathcal{L} }{  \partial (\partial_\mu \partial_\beta  \phi^I )} - \nonumber \\
&  \partial_\gamma \frac{ \partial \mathcal{L}  }{\partial (\partial_\mu \partial_\beta  \partial_\gamma \phi^I )  } + \dots  \bigg\} \partial_\beta \partial^\nu \phi^I  +  \bigg\{   \frac{ \partial \mathcal{L}   }{\partial (\partial_\mu \partial_\beta  \partial_\gamma \phi^I )  } - \dots  \bigg\} \partial_\beta  \partial_\gamma \partial^\nu \phi^I + \dots - \eta^{\mu \nu}  \mathcal{L}. \nonumber \\
\end{align}
It is convenient to introduce another physical object to the current vector
\begin{align}\label{st2}
J^\mu &= i \epsilon \left\{ \frac{ \partial  \mathcal{L}  }{\partial (\partial_\mu \phi^I )  } - \partial_\beta \frac{ \partial  \mathcal{L}  }{  \partial (\partial_\mu \partial_\beta \phi^I ) } + \dots  \right\} \phi^I + \left\{  \frac{ \partial  \mathcal{L} } {  \partial (\partial_\mu \partial_\beta \phi^I )}  + \dots  \right\} \partial^\beta \phi^I + \ldots,
\end{align}
upon substituting \eqref{all:action} into \eqref{st}, we obtain
\begin{equation}\begin{aligned}
T^{\mu\nu} =& \bigg\{ w_o u_\gamma P^{\gamma \mu} ( c_s^2 f_b + ( \gamma_\eta - 2 \gamma_\xi ) b^3 \Delta^{\alpha \beta}_{I J} \partial_\alpha K_{\beta}^K + \gamma_\xi \epsilon^{ \zeta \rho \beta  } \epsilon_{0 L M }    \partial_\zeta \phi^{I} \partial_\rho \phi^{J} K^K_{\beta } - c^2_p b F_y   \\ 
& \times \omega^2 \partial_b \chi^2 + F_b (1-c y^2)  ) + w_o \gamma_\eta  (\Delta^{IJ}_{\alpha \zeta} K^K_{\beta} + K^I_{\alpha} \Delta^{JK}_{\beta \zeta}) \partial^\zeta K^\alpha P^{\mu \beta}  +  F_y \gamma_\chi ( 2 \omega^2 \partial_{\omega^2} \chi   \\ 
& + \chi )( \omega^2 u^\mu  - y_{\beta \theta}  (\dot{K}^\beta - u_\alpha \nabla^\beta K^\alpha) P^{\theta \mu} + \frac{1}{6} \epsilon^{\mu \theta \sigma \gamma} y^{\rho}_{\theta} \partial_\rho \partial_\sigma \phi \partial_\gamma \phi )- \partial_\beta ( z_{IJK}  b^2 \Delta^{IJ}_{\alpha \sigma}  \\
& \times P_K^{\alpha \{ \mu} \delta^{\beta \} \sigma} - 2 c_p^2 F_y\gamma_\chi (1 + 2 \omega \partial_{\omega^2} \chi ) y_{\alpha \theta} P^{\theta \{ \beta} \delta^{ \mu \} \alpha } ) \bigg\} \partial^\nu \phi + \bigg\{ w_o \gamma_\eta \Delta^{IJ}_{\alpha \sigma} P_K^{\alpha \{ \mu} \delta^{\beta \} \sigma} - \\ 
&  \ 2 c^2_p F_y ( 1 + 2 \omega \partial_{\omega^2} \chi ) \gamma_\chi y_{\alpha \theta} P^{\theta \{ \beta} \delta^{ \mu \} \alpha }  \bigg\} \partial_\beta \partial^\nu \phi - \eta^{\mu\nu} \mathcal{L},
\end{aligned}\end{equation}
where we introduce the transport coefficient and thermodynamic derivative
\begin{equation} 
\begin{split}
& c_s^2 = \frac{ f_{bb} b}{f_b}, \quad 
\gamma_\xi \equiv \frac{1}{w_o}  \sum_{   IJK } ( h_{IJK} + \sum_{L M} h_{IJK,LM})
, \quad \chi \equiv \frac{\partial Y^{\mu\nu} }{\partial \omega^{\mu\nu}},
\\
&  c^2_p \equiv  \frac{1}{2} \bigg( 1 + \frac{F_{bb}}{F_b} \bigg), \quad  \gamma_\chi \equiv \frac{\chi^2}{w_o} 
, \quad 
\gamma_\eta \equiv \frac{1}{w_o} \sum_{   IJK  }  z_{IJK}, \quad 
w_o = F_y - f_b b.
\end{split}
\end{equation}
The linearized "equation of motion" describes the macroscopic near-equilibrium system. Now, introducing the identity $u_\mu \Delta^{\mu\nu} = 0$ to fulfill the requirement of projection, we let the equation of motion $\partial_\mu \langle T^{\mu\nu} \rangle = 0$ onto parallel and perpendicular fluid velocity 
\begin{subequations}\begin{align} 
u_\nu \partial_\mu \langle T^{\mu\nu} \rangle = 0, \\ 
\label{trans}
\Delta^\alpha_\nu \partial_\mu \langle T^{\mu\nu} \rangle = 0.
\end{align}\end{subequations}
By following the Taylor expansion around the static equilibrium, we shall express the main out-equilibrium field  $\phi^{iI}(x) = x^{iI} + \pi^{iI} + \frac{1}{2!} \pi \cdot \partial \pi^{iI} + \frac{1}{3!} \pi \cdot \partial (\pi \cdot \partial \pi^{iI}) + \dots$. To calculate the collective modes near-equilibrium limit, we linearize the  equations above around $\phi^{iI} = x^{iI} + \pi^{iI}$ where $x^{iI}$ is the hydrostatic background. The four-velocity is  
\begin{align}
u^\mu = u_0^\mu + \delta u^\mu,     
\end{align}
where $u_0 = 1 + \delta g_{00}/2$ and $ u_\mu \delta u^\mu = 0$, with $\nabla^\mu = \Delta^{\mu \alpha} \partial_\alpha $. One checks by \eqref{Kconser} that
\begin{equation}
u^\mu \simeq \delta^\mu_0 \left( 1 + \frac{1}{2}\dot \pi^2 \right) + \delta^\mu_i \left( \vphantom{\frac{}{}} -\dot \pi^i + \dot \pi\cdot\partial\pi^i  \right). 
\end{equation}
Using the fact that $\omega_{\mu\nu} = \nabla_\mu u_\nu - \nabla_\nu u_\mu $, we similarly derive the same expansion for  
\begin{equation}
\omega^2 \simeq  - (\partial_\mu \dot \pi)\cdot(\partial^\mu \dot \pi) - [\partial\dot \pi \cdot \partial \dot \pi].  
\end{equation}
Following the linearization, the stress tensor and current for the first order are
\begin{eqnarray}
T^{\mu\nu} \supset  && f_b b ( \dot \pi \delta^\mu_0 - \partial_l \dot \pi^j \delta_j^\mu - c_s^2 [\partial \pi] \delta^\mu_l ) \delta^\nu_0 + z_{IJK} ( 3 [\partial \dot \pi ] \delta^{ \mu \nu} - \delta^\mu_I \partial^\nu \dot \pi^I  ) + b h_{IJK,LM} 2 [\partial \dot \pi ] \delta^{\mu \nu} - z_{IJK} \times    \nonumber \\ 
&& (  4 \partial_l [\partial \pi] \delta^\mu_l \delta^\nu_0 - \partial^\mu \partial^\nu \pi + \partial^\mu \dot \pi^J  \delta^\nu_J  - \partial_l \partial^m \pi \delta^\mu_l \delta^\nu_m + 2 [\partial \dot \pi] \delta^{\mu \nu} ) + \chi^2 ( \ddot \pi^l \delta^\mu_l + \partial^2 \dot \pi^l \delta^\mu_l + \partial^\mu \ddot \pi  +  \nonumber \\ 
&& \partial^\mu [ \partial \ddot \pi]) \delta^\nu_0 - \eta^{\mu\nu} [\partial \pi], 
% J^l_\mu  \supset  && \ f_b b ( \delta_\mu^0 \dot \pi^l  - \partial_l \pi^j \delta^j_\mu - c_s^2 [\partial \pi ] \delta^j_\mu  ) + z_{IJK} ( \partial_l \dot \pi^j \delta^j_\mu  - 4 \delta_\mu^0 \partial_l [\partial \pi] + \delta^0_\mu \partial_l [\partial \pi] ) + 2 b_o^2 h_{IJK,LM}  \times    \nonumber \\
% &&  ( [\partial \dot \pi] \delta_\mu^l + \partial_l [\partial \pi ] \delta_\mu^0 ) - \chi^2 (\partial_\mu \ddot \pi^l + \dddot \pi^l + \partial_l \ddot \pi^j \delta^j_\mu + \partial_l [\partial \dot \pi ] \delta_\mu^0 ). 
\end{eqnarray}
The Green-Kubo formalism for variational principle points out $g^{\mu\nu}$ and $\omega^{\mu\nu}$ as the independent background. To avoid self-interaction and second order terms, we use $g^{\mu\nu} = \eta^{\mu\nu} + h^{\mu\nu}$
through the covariant derivative
% $\omega^{\mu\nu} = (0,0) + \delta \omega^{\mu\nu}$ and
\begin{equation}
\nabla_\mu u^\nu = \partial_\mu u^\nu + \frac{1}{2} \eta^{\nu \beta} (\partial_\mu h_{\beta \rho} + \partial_\rho h_{\beta \mu} - \partial_\beta h_{\mu \rho})u^\rho_0.   
\end{equation}
Choosing the metric direction parallel to the external vortex field $\delta g^{\mu\nu} = \delta g^{\mu\nu} (t, x_3)$. We employ the definition in \eqref{eq:GR-sources} to evaluate the retarded functions $ \delta \phi^{iI} (\omega, \textbf{k}) = \int d \omega d \textbf{k}^3 e^{i \omega t - i \textbf{k}x } \delta \phi^{iI} (t,x) $. The correlation functions for dissipative spin hydrodynamics are 
\begin{eqnarray}
G_{T^{xz},T^{xz}} &=& \frac{  (\omega^2 + \omega \textbf{k} ) + \chi^2   (\omega^5 - 2 \omega^3 \textbf{k}^2 ) + i z_{IJK} \chi^2 (2 \omega^5 - \omega^4 \textbf{k} + \omega^2 \textbf{k}^3 ) + z_{IJK}^2 (\omega \textbf{k}^4 - 4 \omega^2 \textbf{k}^3     }{ \omega^2 + \textbf{k}^2 - i z_{IJK} \omega \textbf{k}^2 - 2 i h_{IJK,LM} \omega \textbf{k}^2 - 2 \chi^2 \omega^4 } \nonumber\\
&& \frac{ - 3 \omega^3 \textbf{k}^2 ) + i   z_{IJK} \omega^3 - 4 \chi^4 (\omega^6 - \omega^4 \textbf{k}^2)}{},  \\
G_{T^{xy},T^{xy}} &=& \frac{ \omega \textbf{k} - h_{IJK,LM}    (\omega^4 - c_s^2 \omega^2 \textbf{k}^2 ) + h_{IJK,LM} \chi^2 (\omega^4 \textbf{k}^2 - \omega^3 \textbf{k}^3) + z_{IJK} \chi^2 (\omega^6 - \omega^4 \textbf{k}^2)}{  \omega^2 - c_s^2 \textbf{k}^2 - i z_{IJK} (2 \omega^2 \textbf{k} + 3 \omega \textbf{k}^2) - 2 i h_{IJK,LM} \omega \textbf{k}^2 - 2 \chi^2 (\omega^4 - \textbf{k}^2 \omega^2)}  \nonumber\\ 
&& \frac{   - 3 z_{IJK}^2 (\omega^4 \textbf{k} + \omega^3 \textbf{k}^2) - 2 z_{IJK}^{'2} \omega^2 \textbf{k}^2 + \chi^4 ( 2 \omega^7 + 2 \omega^6 \textbf{k} - \omega^4 \textbf{k}^3 )}{},\label{eq:state-alice-antirob}\\ 
G_{T^{0z},T^{0z}}&=&\frac{  z_{IJK} (\omega^4 - 2 \textbf{k}^4 ) - 2   \chi^2 ( \omega^4 + \omega^2 \textbf{k}^2 ) + z_{IJK}^2 (\omega^4 - 4 \omega^2 \textbf{k}^2 + \omega^3 \textbf{k} - 4 \omega \textbf{k}^3 ) + 8 h_{IJK,LM} }{  \omega^2 - i z_{IJK} \omega \textbf{k}^2 - 2 i h_{IJK,LM} \omega \textbf{k}^2 - 2 \chi^2 \omega^4} \nonumber\\ 
&& \frac{ \times \chi^2 \omega^2 \textbf{k}^2 + 2 \chi^4 ( \omega^5 \textbf{k} - \omega^4 \textbf{k}^2 - \omega^2 \textbf{k}^4 )}{}, \label{eq:state-rob-antirob}\\
G_{T^{0x},T^{0x}} &=& \frac{   ( 4 z_{IJK} + h_{IJK,LM} )\omega^2 \textbf{k}^2 + \chi^2 ( 2 \omega^5 - \omega^3 \textbf{k}^2 ) + z_{IJK} \chi^2 ( 2 \omega^5 \textbf{k} + 5 \omega^4 \textbf{k}^2 - 3 \omega^2 \textbf{k}^4 ) + }{\omega^2 - i z_{IJK} \omega \textbf{k}^2 - 2 i h_{IJK,LM} \omega \textbf{k}^2 - 2 \chi^2 \omega^4} \nonumber\\ 
&& \frac{  2 h_{IJK,LM} \chi^2 ( \omega^4 \textbf{k}^2 + \omega^2 \textbf{k}^4 ) - 4 \chi^4 ( \omega^5 \textbf{k} - \omega^3 \textbf{k}^3 )}{}, \\
G_{T^{00},T^{00}} &=& \frac{ ( \omega \textbf{k} +  \textbf{k}^2 ) - 2   z_{IJK} (\omega^3 \textbf{k} + 2 \omega^2 \textbf{k}^2 + 5 \omega \textbf{k}^3) +   h_{IJK,LM} ( 4\omega^2 \textbf{k}^2 + \omega \textbf{k}^3 ) + \chi^2 (\omega^3 \textbf{k} }{ \omega^2 - c_s^2 \textbf{k}^2 - i z_{IJK} (2 \omega^2 \textbf{k} + 3 \omega \textbf{k}^2) - 2 i h_{IJK,LM} \omega \textbf{k}^2 - 2 \chi^2 (\omega^4 - \textbf{k}^2 \omega^2)} \nonumber\\ 
&& \frac{ + \omega^2 \textbf{k}^2  ) + z_{IJK}^2 (5 \omega \textbf{k}^3 + \omega^2 \textbf{k}^2 )  + 2  \chi^2 (z_{IJK} + h_{IJK,LM}) ( \omega^4 \textbf{k}^2 - \omega^2 \textbf{k}^4 + \omega^3 \textbf{k}^3 ) -   }{} \nonumber\\
&& \frac{ \chi^4 ( \omega^5 \textbf{k} + \omega^4 \textbf{k}^2 - \omega^3 \textbf{k}^3 )}{}.
\end{eqnarray}
It is straightforward the above calculation for conserved current 
\begin{align}
G_{J^z,\omega^{xy}} &= \frac{ 4 i \omega^2 \textbf{k} + 4 \chi^2 z_{IJK} \omega^3 \textbf{k} }{  (\omega^2 - c_s^2 \textbf{k}^2) - i  \chi^2 (\omega^4 - \omega^2 \textbf{k}^2) } - 2 \chi^2 \omega^2, \\
G_{J^x,\omega^{xy}} &= \frac{  2 z_{IJK}^2 \omega^2 \textbf{k}^2 }{  (\omega^2 - c_s^2 \textbf{k}^2) - i  \chi^2 (\omega^4 - \omega^2 \textbf{k}^2)} - z_{IJK} \omega \textbf{k},  \\
G_{J^0,T^{00}} &= \frac{ \omega^2 \textbf{k} + 3 z_{IJK}^2 \omega \textbf{k}^4 - \chi^4 ( \omega^5 \textbf{k} + \omega^3 \textbf{k}^3 )}{  (\omega^2 - c_s^2 \textbf{k}^2) + 8 i z_{IJK} \omega \textbf{k}^2 - \chi^2 (\omega^4 - \omega^2 \textbf{k}^2 )  } + z_{IJK} ( \textbf{k}^2 - \omega k) - 2 \chi^2 (\omega^2 - \textbf{k}^2), \\
G_{J^z,T^{xy}} &= \frac{  z_{IJK} \omega \textbf{k}^2 + 3 z_{IJK}^2 \omega \textbf{k}^4 + z_{IJK} \chi^2 (\omega^2 \textbf{k}^3 ) - \chi^4 ( \omega^4 \textbf{k}^2 + \omega^3 \textbf{k}^3 )}{  (\omega^2 - c_s^2 \textbf{k}^2) + 8 i z_{IJK} \omega \textbf{k}^2 - \chi^2 (\omega^4 - \omega^2 \textbf{k}^2 )} + iz_{IJK} ( \textbf{k}^2 - \omega \textbf{k})\textbf{k}    + \chi^2 \omega^2. 
\end{align}

% Similar to \eqref{st2}, we get
% \begin{equation}\begin{aligned} 
% J^\mu &= \bigg\{   u_\gamma P^{\gamma \mu} \big\{ f_b(b) + ( z_{IJK,LM} (b^2) + z_{ijk} (b^2) ) b^3 \Delta^{\alpha i \beta j} \partial_\alpha K^k_\beta + z_{ijk}(b^2) \epsilon^{\mu \chi \rho \beta } \epsilon_{0IMN}  \partial_t \partial_\chi \phi^{Mi} \partial_\rho \phi^{Nj} K^{\beta k}    \\ 
% & + f_b (1-c y^2) - c b F_y \omega^2 \partial_b \xi^2 + z_{ijk} (b^2) (\Delta^{ij}_{\alpha \gamma} K^k_{\beta} + \Delta^{kj}_{\beta \gamma} K^i_{\alpha} ) - 2 c F_y ( 2 \omega^2 \partial_{\omega^2} \chi + \chi ) ( - \chi \omega^2 u_\theta + y_{\beta \theta} (\dot{K}^\beta - u_\alpha \nabla^\beta K^\alpha) P^{\theta \mu}  \\ 
% & + \frac{1}{6} \epsilon^{\mu \theta \sigma \gamma}  y^{\rho}_{\theta} \partial_\rho \partial_\sigma \phi^I \partial_\gamma \phi^K ) - \partial_\beta ( z_{ijk} (b^2) b^2 \Delta^{ij}_{\alpha \sigma} P^{\alpha k \{ \mu} \delta^{\beta \} \sigma} - 2 c F_y (\chi + 2 \chi \omega \partial_{\omega^2} \chi )y_{\alpha \theta} P^{\theta \{ \beta} \delta^{ \mu \} \alpha } )\bigg\} \phi^I
% \end{aligned}\end{equation}

\section{Further properties of the two-point function in hydrodynamics}\label{twop}

We show the propagator of shear and bulk in the effective field theory. We can physically distinguish the local couplings between ideal \eqref{ideal:} and viscous fluid \cref{eqsshear,eqsbulk}. The diagrams for the 2-point correlator function $ \langle \delta K^0 , \delta K^0 \rangle $ at $\mathcal{O} (w_o^{-1})$ for longitudinal and transverse excitations are
\begin{equation}\begin{aligned}\label{free:ns}
D_{ij}^{(0)} (\omega,\textbf{k}) \,\, &= \frac{1}{w_o} \frac{iL_{ij}}{\omega^2-c_s^2 \textbf{k}^2 + i \gamma_\eta (z)\omega \textbf{k}^2 }  +  \frac{1}{w_o} \frac{iT_{ij}}{\omega + i \gamma_\eta (z) \textbf{k}^2 },
\end{aligned}\end{equation}
where the transverse and longitudinal projectors are $T_{ij} = ( \delta_{ij} - k_i k_j / \textbf{k}^2 )$ and $L_{ij} = k_i k_j / \textbf{k}^2 $, respectively. One obtains for the longitudinal three different propagators which involves the spatial and time mixing
\begin{equation}\begin{aligned}
\langle [ \partial \pi ] [ \partial \pi ] \rangle = \frac{1}{w_o} & \frac{i  \textbf{k}^2 }{\omega^2-c_s^2 \textbf{k}^2 + i \gamma_\eta (z)\omega \textbf{k}^2},  \qquad \qquad \qquad \qquad  \langle \dot \pi_l  [ \partial \pi ] \rangle  = \frac{1}{w_o}  \frac{i \omega \textbf{k}_l }{\omega^2-c_s^2 \textbf{k}^2 + i \gamma_\eta (z)\omega \textbf{k}^2}, \\
& \quad \quad \qquad \qquad \langle \dot \pi_i  \dot \pi_j \rangle = \frac{1}{w_o} \frac{i  \delta_{ij} \omega^2 - c_s^2 \textbf{k}^2 T_{ij}}{\omega^2-c_s^2 \textbf{k}^2 + i \gamma_\eta (z)\omega \textbf{k}^2}.
\end{aligned}\end{equation}
They are useful for the evaluation of Feynman diagrams. The 2-point correlator function $ \langle \delta K^0 , \delta K^0 \rangle $ from the polarization Lagrangian \eqref{eqspol} at $\mathcal{O} (w_o^{-1})$ for longitudinal and transverse propagators obey 

\begin{equation}\begin{aligned}\label{free:pl}
H_{ij}^{(0)} (\omega,\textbf{k}) \,\, &= \frac{1}{w_o} \frac{iL_{ij}}{\omega^2 - c_p^2 \textbf{k}^2 - \chi^{-2}  } + \frac{1}{w_o} \frac{i T_{ij} (\omega^2 - \textbf{k}^2)}{(\omega^2 - \textbf{k}^2) - \chi^{-2}}.
\end{aligned}\end{equation}

\end{document}